\documentclass[traditabstract]{aa} 

\usepackage{txfonts}
\usepackage{graphicx}
\usepackage{natbib}

\newcommand{\phiorb}{$\phi_{98}$}
\newcommand{\phispin}{$\phi_{67}$}

\newcommand{\oex}{\mbox{\object{EX\,Hya}}}

\newcommand{\hbet}{H$\beta$}
\newcommand{\hgam}{H$\gamma$}
\newcommand{\hdel}{H$\delta$}

\newcommand{\hei}{\ion{He}{i}}
\newcommand{\heii}{\ion{He}{ii}}

\newcommand{\caii}{\ion{Ca}{ii}$\lambda 8498$}
\newcommand{\fei}{\ion{Fe}{i}$\lambda 4957$}
\newcommand{\porb}{$P_\mathrm{orb}$}
\newcommand{\pspin}{$P_\mathrm{spin}$}

\newcommand{\gc}{G\,cm$^{3}$}

\newcommand{\gcms}{g\,cm$^{2}$s$^{-2}$}
\newcommand{\kms}{km\,s$^{-1}$}

\newcommand{\ergs}{erg\,cm$^{-2}$s$^{-1}$}
\newcommand{\ergsa}{erg\,cm$^{-2}$s$^{-1}$\AA$^{-1}$}

\newcommand{\ten}[2]{#1\,\times\,10^{#2}}

\newcommand{\mwd}{$M_\mathrm{wd}$}

\newcommand{\reins}{$R_1$}
\newcommand{\vterm}{$\upsilon_\mathrm{term}$}
\newcommand{\vrad}{$\upsilon_\mathrm{rad}$}
\newcommand{\vkep}{$\upsilon_\mathrm{kep}$}

\newcommand{\rin}{$r_\mathrm{in}$}

\newcommand{\tmag}{$t_\mathrm{mag}$}
\newcommand{\rcirc}{$r_\mathrm{circ}$}
\newcommand{\rco}{$r_\mathrm{cor}$}

\newcommand{\hsh}{$h_\mathrm{sh}$}

\newcommand{\msun}{$M_\odot$}
\newcommand{\msunyr}{$M_{\odot}$yr$^{-1}$}
\newcommand{\oc}{$O\!-\!C$}

\begin{document}

\title{\mbox{High-resolution spectroscopy of the intermediate polar EX
    Hydrae}: II. The inner disk radius \thanks{Based on observations collected with the ESO Very Large
    Telescope, Paranal, Chile, in program 072.D--0621(A).}}

\author{K.~Beuermann\inst{1} \and K.~Reinsch\inst{1}}
 
\institute{ 
  Institut f\"ur Astrophysik und Geophysik, Georg-August-Universit\"at,
  Friedrich-Hund-Platz 1, D-37077 G\"ottingen, Germany }

\date{Received 21 December 2023 ; accepted 25 January 2024}

\abstract{\oex\ is one of the best studied, but still enigmatic
  intermediate polars.  We present phase-resolved blue VLT/UVES
  high-resolution ($\lambda/\Delta \lambda \simeq 16.000$) spectra of
  \oex\ taken in January 2004. Our analysis involves a unique
  decomposition of the Balmer line profiles into the spin-modulated
  line wings that represent streaming motions in the magnetosphere and
  the orbital-phase modulated line core that represents the accretion
  disk.  Spectral analysis and tomography show that the division line
  between the two is solidly located at
  $\mid\!\upsilon_\mathrm{rad}\!\mid\,\simeq\!1200$\,\kms, defining
  the inner edge of the accretion disk at
  $r_\mathrm{in}\!\simeq\!\ten{7}{9}$\,cm or $\sim\!10$\,\reins\ (WD
  radii). This large central hole allows an unimpeded view of the tall
  accretion curtain at the lower pole with a shock height up to
  \hsh$\sim\!1$\,\reins\ that is required by X-ray and optical
  observations. Our results contradict models that advocate a small
  magnetosphere and a small inner disk hole.  Equating $r_\mathrm{in}$
  with the magnetospheric radius in the orbital plane allows us to
  derive a magnetic moment of the WD of
  $\mu_1\!\simeq\!\ten{1.3}{32}$\,\gc\ and a surface field strength
  $B_1\!\sim\!0.35$\,MG. Given a polar field strength
  $B_\mathrm{p}\!\protect\la\!1.0$\,MG, optical circular polarization
  is not expected.  With an accretion rate
  $\dot M\!=\!\ten{3.9}{-11}$\,\msunyr, the accretion torque is
  $G_\mathrm{acc}\!\simeq\!\ten{2.2}{33}$\,\gcms. The magnetostatic
  torque is of similar magnitude, suggesting that \oex\ is not far
  from being synchronized.  We measured the orbital radial-velocity
  amplitude of the WD, $K_1=58.7\pm3.9$\,\kms, and found a
  spin-dependent velocity modulation as well.  The former is in
  perfect agreement with the mean velocity amplitude obtained by other
  researchers, confirming the published component masses
  $M_1\simeq0.79$\msun\ and $M_2\simeq0.11$\,\msun.}
  
   
\titlerunning{Inner disk radius of \oex}
\authorrunning{K.~Beuermann \& K.~Reinsch}

\maketitle


\section{Introduction}
\label{sec:intro}

\oex\ is the best studied member of the intermediate polar (IP)
subclass of cataclysmic variables (CVs) that has about 150 confirmed
members and candidates (see Koji Mukai's IP catalog\footnote{https://asd.gsfc.nasa.gov/Koji.Mukai/iphome/catalog/alpha.html}).
With an orbital period of 98.26\,min, it is one of the few IPs below
the period gap. \oex\ has been intensely studied over the last 60
years, facilitated by its small distance of $d=56.77\pm0.05$\,pc
\citep{bailerjonesetal21} and  very low interstellar column density.
Nevertheless, the fundamental IP parameters, such  as the magnetic moment of the
white dwarf (WD), remain uncertain. While there is agreement that \oex\
accretes from a disk via the magnetosphere of the WD, to date it has not been
possible  to reliably measure the radius \rin\ of the boundary
between disk and magnetosphere, a key ingredient for studies of the
accretion geometry. The WD in \oex\ is a slow rotator with
\pspin$=2\pi/\omega=67.03$\,min and the corotation radius \rco\
approximately equals the Roche radius of the WD, exceeding the
circularization radius \rcirc\ by a factor of about three. If \oex\
harbors a standard disk that extends inward to \rin$<r_\mathrm{circ}$,
the Keplerian speed at the boundary between disk and magnetosphere
substantially exceeds the rotation speed
$\upsilon_\mathrm{in}= \omega\,$\rin\ of the field. Accretion then
occurs far from spin equilibrium, spinning up the WD on a  timescale
of $\ten{5}{6}$\,yr \citep{maucheetal09,andronovbreus13}.  The
uncertainties with regard to  the inner disk radius and the magnetic moment of
the WD have led to divergent models of the accretion geometry of \oex.

The large-magnetosphere model is based on the suggestion of
\citet{kingwynn99} that accretion in \oex\ may proceed near spin
equilibrium, provided the magnetic field is strong and governs the
accretion flow close to $L_1$. A ring is formed in the outer Roche
lobe of the WD, from which it accretes over 360\degr\ in azimuth,
preventing the system from becoming a polar
\citep{nortonetal04,nortonetal08}. Other authors have discussed their data
in the framework of this model \citep{belleetal02,mhlahloetal07}.  The
lack of observable circular polarization \citep{buttersetal09} argues
against it.

\begin{table}[b]
  \caption{Journal of the VLT/UVES observations of EX Hya in January
    2004. The blue spectra cover the wavelength range $3750-4980$\AA.}
\begin{tabular}{llccc} 
\hline \hline \noalign{\medskip}
Date & Target & UT & Number  & Exposure (s)\\ 
\noalign{\medskip} \hline
\noalign{\medskip}
Jan 23, 2004 & EX Hya & 5:42 -- 8:17 & 26 & 300 \\[0.5ex]
Jan 26, 2004 & EX Hya & 6:25 -- 8:38 & 25 & 300 \\[0.5ex]
\noalign{\medskip} \hline      
                                 
\end{tabular}
\label{tab:obslog}
\end{table}

The extended wings of the Balmer lines in \oex\ were originally
assigned to the inner accretion disk of what was then considered a
dwarf nova \citep{breysachervogt80,cowleyetal81,gilliland82}.  This
was an obvious choice before the magnetic nature of the WD was finally
recognized
\citep{vogtetal80,kruszewskietal81,warnermcgraw81,sterkenetal83,cordovaetal85}.
Since the flux in the line wings varies with the spin period, they
were subsequently associated with streaming motions in the
magnetosphere \citep{cordovaetal85,hellieretal87,kaitchucketal87},
although several authors continued to favor a disk origin of part of
the line wings  \citep[e.g.,][]{belleetal03,echevarriaetal16}. The
small-magnetosphere model was formally introduced by
\citet{revnivtsevetal09}, who suggested that an observed break in the
power spectra of the X-ray and optical intensity fluctuations of an IP
may indicate the orbital frequency at the inner edge of its accretion
disk. This model was employed to \oex\ by \citet{revnivtsevetal11},
\citet{semenaetal14}, and \citet[][and references
therein]{suleimanovetal16,suleimanovetal19} and predicts inner disk
radii between \rin$\,=\!2$ and 4\,\reins, with \reins\ the WD radius.
The implication is that Balmer line emission up to velocities between
1800 and 2600\,\kms\ originates from the inner disk.
\citet{lunaetal18} showed this concept to be inconsistent with X-ray
results, rendering the size of the inner hole of the disk in \oex\ a
controversial issue that is, at the same time, fundamental to the
discussion of its accretion geometry.

In this paper, we present a unique decomposition of the
emission line profiles into the spin-modulated line wings and the
orbital-phase dependent line core. The latter includes emissions from
the disk, the S-wave, and the secondary star, and is loosely referred
to as the disk component. Its tomo\-graphy allows a direct measurement
of the terminal velocity of the disk component, which translates into
the inner radius of the accretion disk. In addition, we find a slight
asymmetry or ellipticity of the motion at \rin\ that is also predicted
in hydrodynamic model calculations \citep{bisikaloetal20,kingwynn99}.

\section{Observations}
\label{sec:obs}

\begin{table}[b]
  \caption{Sine fits to narrow-line radial velocities using the
      \citet{helliersproats92} (HS92) linear orbital ephemeris as
      reference. The columns give the NIST wavelength, the velocity
      amplitude, the observed phase offset of the zero crossing from
      HS92 (see text), the systemic velocity, and $\chi^2$ for the
      quoted number of degrees of freedom.}
\begin{tabular}{l@{\hspace{4mm}}c@{\hspace{3mm}}c@{\hspace{3mm}}c@{\hspace{3mm}}c@{\hspace{2mm}}c} 
\hline \hline \noalign{\smallskip}
  Ion & $\lambda _\mathrm{NIST}$ & $K_2'$  & $\phi _0$ & $\gamma $ & $\chi^2$/d.o.f. \\
      & ($\AA $) & (\kms )  &   & (\kms ) & \\
\noalign{\smallskip} \hline
\noalign{\medskip}
  \ion{Ca}{ii} & 8498.020 & $357.2\pm5.0$ & 0.0091(26) & $-57.8\pm3.8$ & ~6.5/7 \\
  \ion{Fe}{i}  & 4957.596 & $346.3\pm4.2$ & 0.0109(22) & $-59.5\pm3.2$ & ~6.5/7 \\
  \ion{Si}{i}  & 3905.525 & $346.0\pm5.0$ & 0.0070(28) & $-54.3\pm4.2$ & ~8.5/8 \\[0.5ex]
\noalign{\smallskip} \hline      
\end{tabular}
\label{tab:narrow}
\end{table}

\subsection{Basic information}
\oex\ was observed in service mode with the Ultraviolet and Visual
Echelle Spectrograph (UVES) at the Kueyen (UT2) unit of the ESO Very
Large Telescope (VLT), Paranal/Chile, on January 23 and 26, 2004,
continuously for 2.58 and 2.50 h, respectively. A total of 26 and 25
spectra were taken. Exposure times were 300\,s with dead times of
typically 60\,s between exposures, due to readout and overhead. The
resulting orbital phase resolution is
$\Delta \phi_{98}\!\simeq\!0.061$ and the spin-phase resolution
$\Delta \phi_{67}\!\simeq\!0.090$. Spectra in the blue and red arms
were obtained simultaneously;  the red spectra were
extensively discussed by \citet[][Paper I, hereafter
BR08]{beuermannreinsch08}. Table~1 contains a log of the
observations. We adopted the ESO UVES pipeline reduction that provides
flat fielding, performs a number of standard corrections, extracts the
individual echelle orders, and collects them into a combined
spectrum. The flux calibration, except for the seeing correction
considered below, is also part of the pipeline reduction. The
wavelength calibration is derived from spectra taken with a ThAr
lamp. The blue spectra cover the wavelength range 3760--4980\AA\ with
a pixel size of 0.030\,\AA.  With a slit width of 1.0\arcsec, the full width at half maximum (FWHM)
resolution is 0.175\AA\ and the spectral resolution is
$\lambda/\Delta \lambda\!\simeq\!27\,000$ at \hbet. To improve the S/N, we rebinned the data into 0.300\,\AA\ bins, giving an effective
resolution of $\lambda/\Delta\!\lambda\!\simeq\!16\,000$ or
$\Delta \upsilon\!=\!18.5$\,\kms\ at \hbet. All times are quoted as
UTC and were converted to BJD(TDB) using the tool provided by
\citet{eastmanetal10}.\footnote{
  http://astroutils.astronomy.ohio-state.edu/time/ converts UTC to
  Barycentric Julian Date in Barycentric Dynamical Time BJD(TDB). }

\begin{figure}[t]
\includegraphics[height=60.0mm,angle=0,clip]{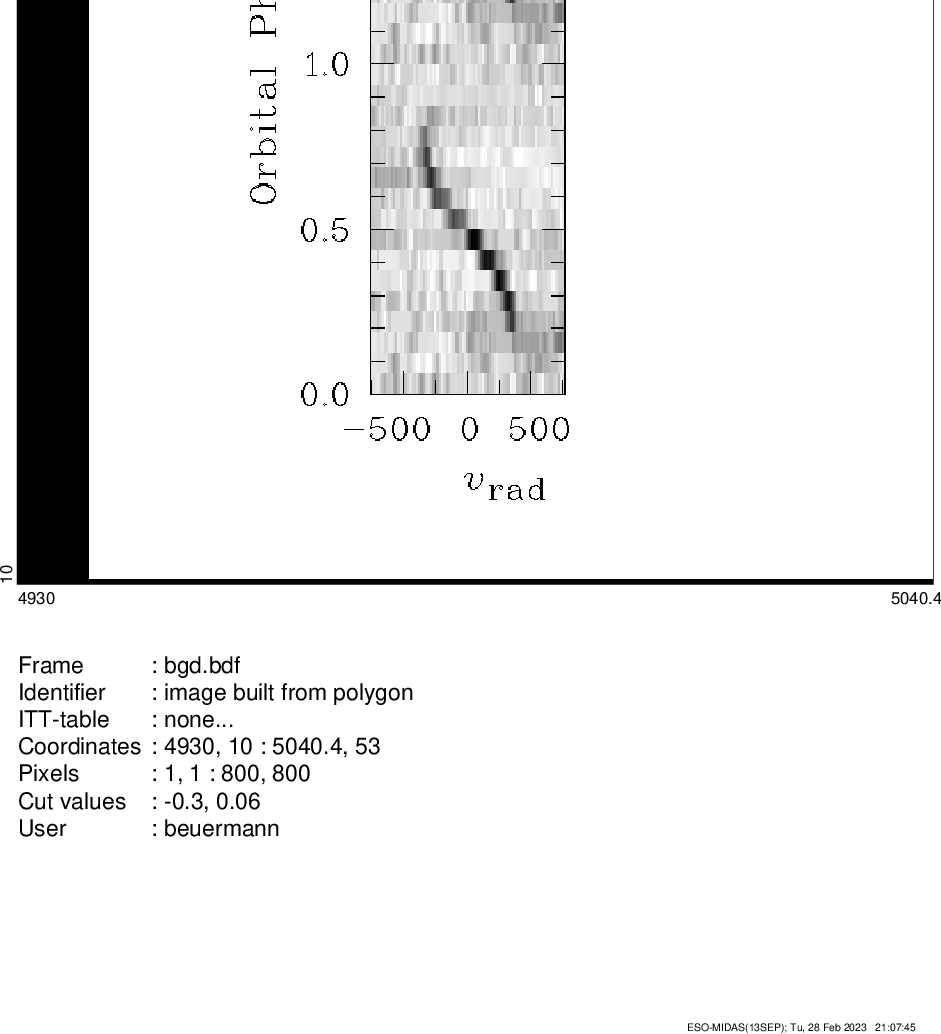}
\includegraphics[height=60.0mm,angle=0,clip]{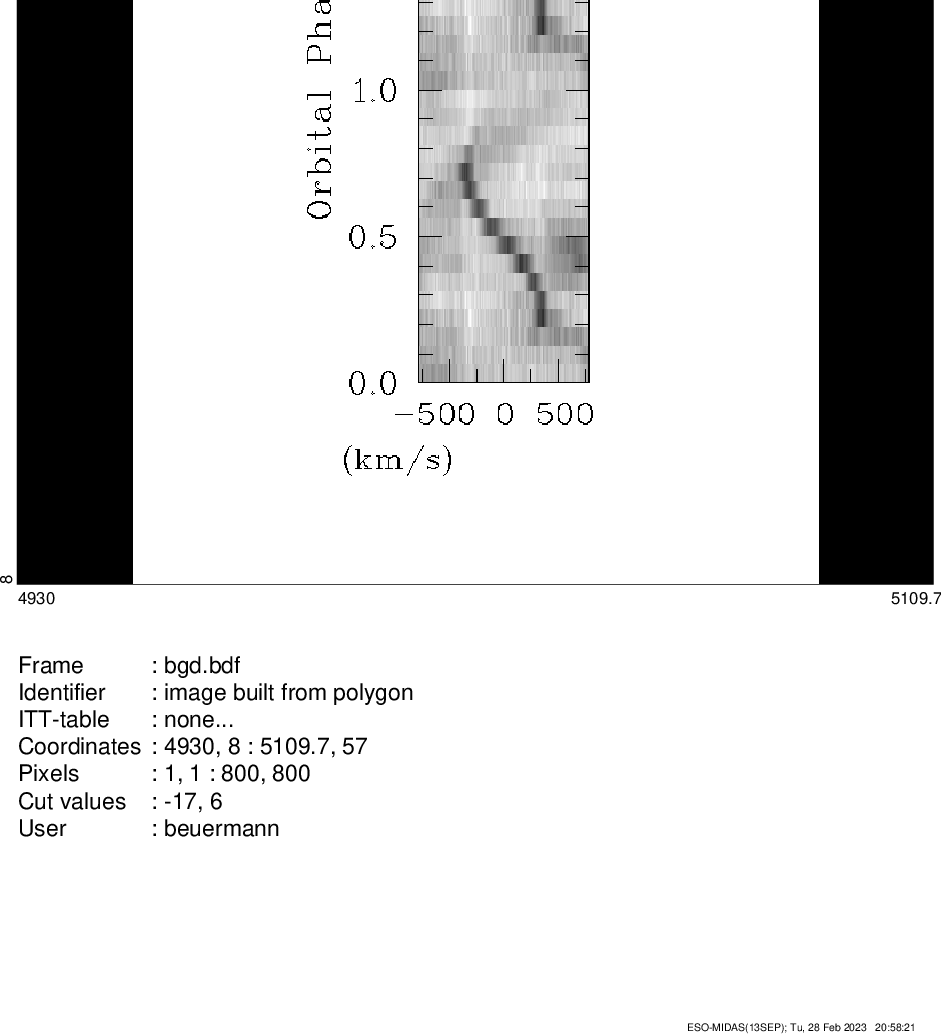}
\hfill
\begin{minipage}[b]{38.0mm}
\includegraphics[width=36mm,angle=0,clip]{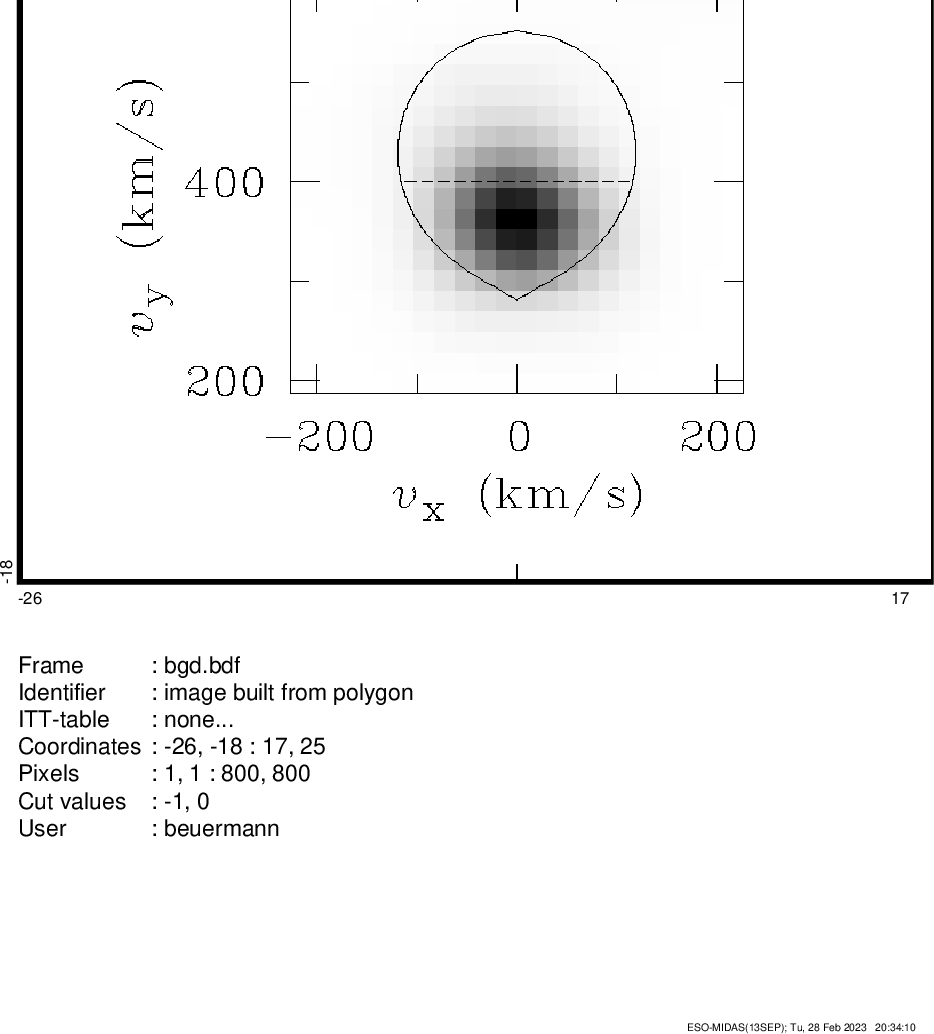}

\includegraphics[width=36mm,angle=0,clip]{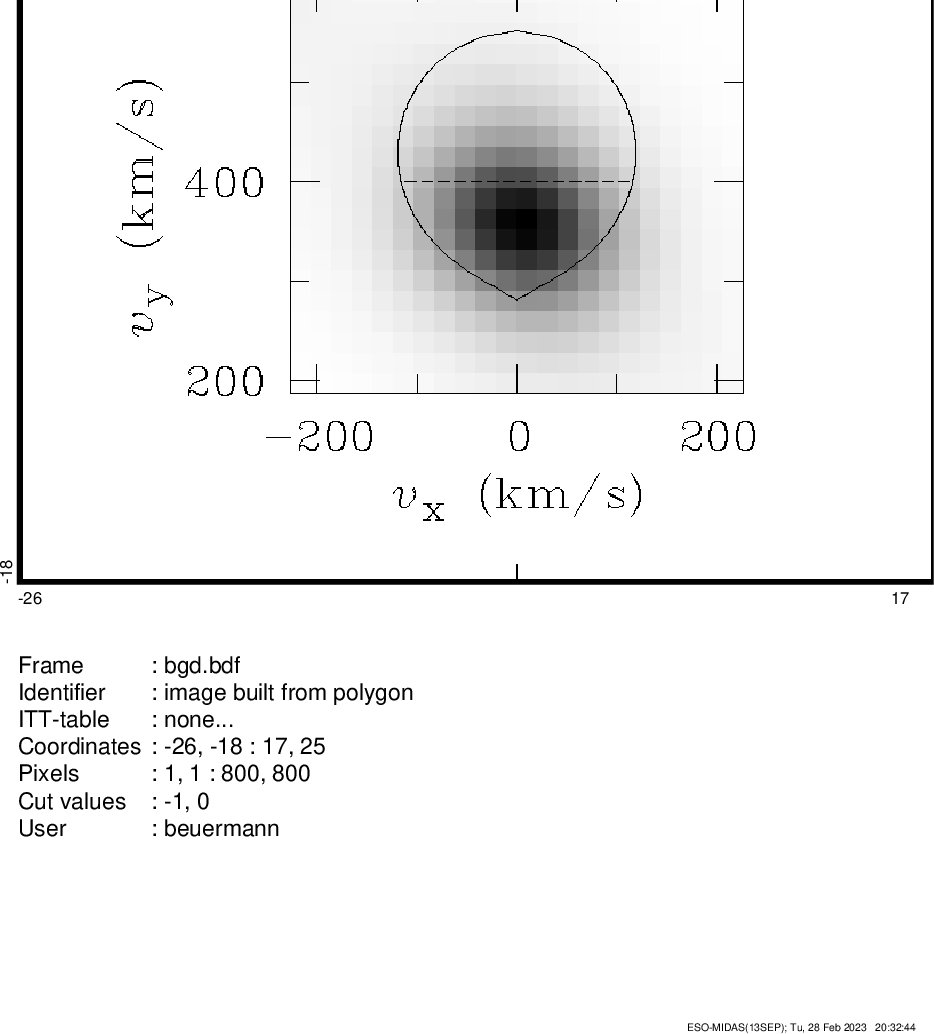}
\end{minipage}
\caption{
  VLT/UVES narrow emission line spectra of FeI4957 and Ca8498 observed
  on 23 and 26 January 2004.  Left: Grayscale representations of the
  trailed spectra. Spectral flux increases from white to black. For
  convenience, the data are shown twice.  Right: Corresponding
  tomograms with the Roche lobe contour of the secondary star based on
  the system parameters of Sect.~\ref{sec:system} demonstrate the
  origin from the irradiated face of the secondary star. The relative
  intensity goes from white (0.0) to black (1.0).  The origin of the
  coordinate system is at the center of gravity.}
\label{fig:tomosec}
\end{figure}

\subsection{Wavelength calibration and phase convention}

The wavelength calibration of the red spectra was based on telluric
molecular lines and was shown to be accurate to 1\,\kms\ (BR08). These
lines are not present in the blue spectra, which contain, instead,
numerous narrow metal emission lines from the illuminated face of the
secondary star (BR08, their Sect.~4.2) that allow a check on the
wavelength calibration. Trailed spectra of \ion{Ca}{ii}$\lambda8498$
and \ion{Fe}{i}$\lambda 4957$ from the observations on 23 and 26
January 2004 are displayed in Fig.~\ref{fig:tomosec}.  We measured the
radial velocities of these lines and of \ion{Si}{i}$\lambda 3905$, by
fitting single Gaussians. The velocities are well represented by
sinusoids
\begin{equation}
    \upsilon_\mathrm{rad} = K_2'\,\mathrm{sin}[2\pi(\phi_\mathrm{HS}+\phi_0)]+\gamma,
\end{equation}
where $\phi_\mathrm{HS}$ refers to the eclipse ephemeris of
\citet{helliersproats92} against which we measure phase offsets. The
fit parameters $K_2'$, $\phi_0$, and $\gamma$ are listed in
Table~\ref{tab:narrow}.  The two rather faint blue lines have
$\gamma=-57\pm 3$\,\kms\ relative to the NIST
wavelengths,\footnote{The US Department of Commerce National
  Institute of Standards and Technology:
  https://www.nist.gov/pml/atomic-spectra-database} which is
indistinguishable from the mean $\gamma=-58.2\pm 1.0$\,\kms\ of the
brighter red lines (BR08). We adopt the latter value also for the blue
spectra and conclude that the wavelength calibrations in both arms are
similarly accurate.

\begin{figure}[t]
\hspace{-2mm}
\includegraphics[width=52.0mm,angle=270,clip]{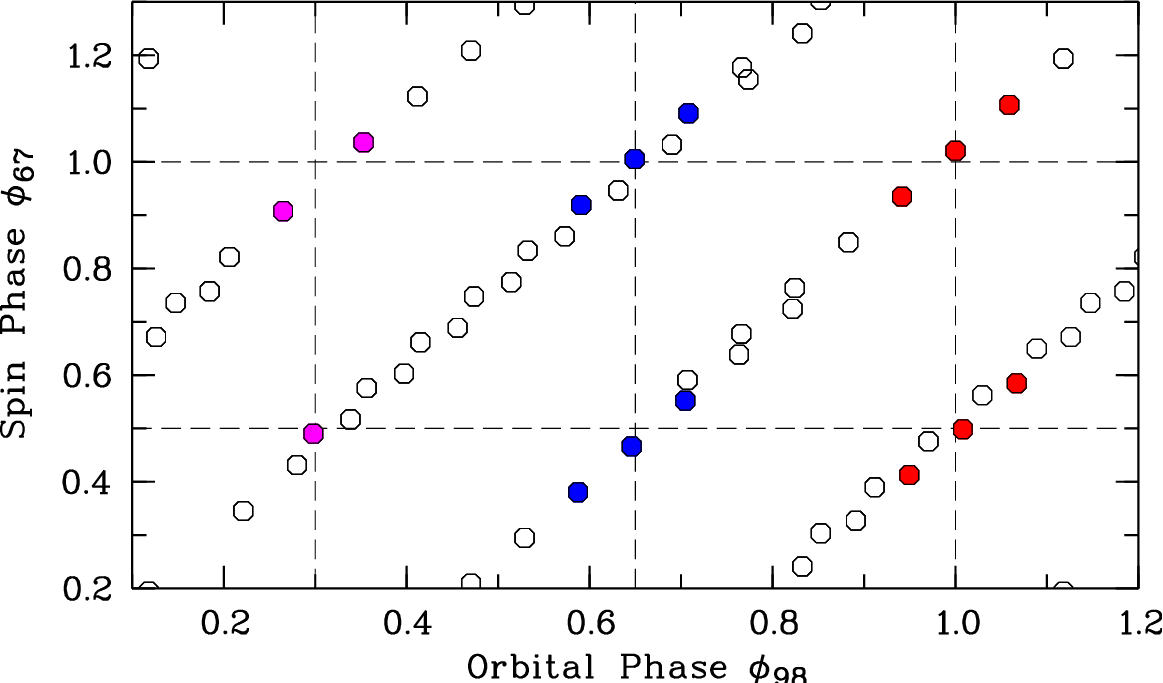}
\caption{
  Phase-space coverage of our 51 exposures in the
  \phispin\--\phiorb\ plane. The spectra selected for the spectral
  decomposition at \phiorb$\,\simeq 0.30, 0.65$, and
  1.00 are marked by magenta, blue, and red dots respectively.}
\label{fig:phase}
\end{figure}

\begin{figure}[t]
\includegraphics[width=52.0mm,angle=270,clip]{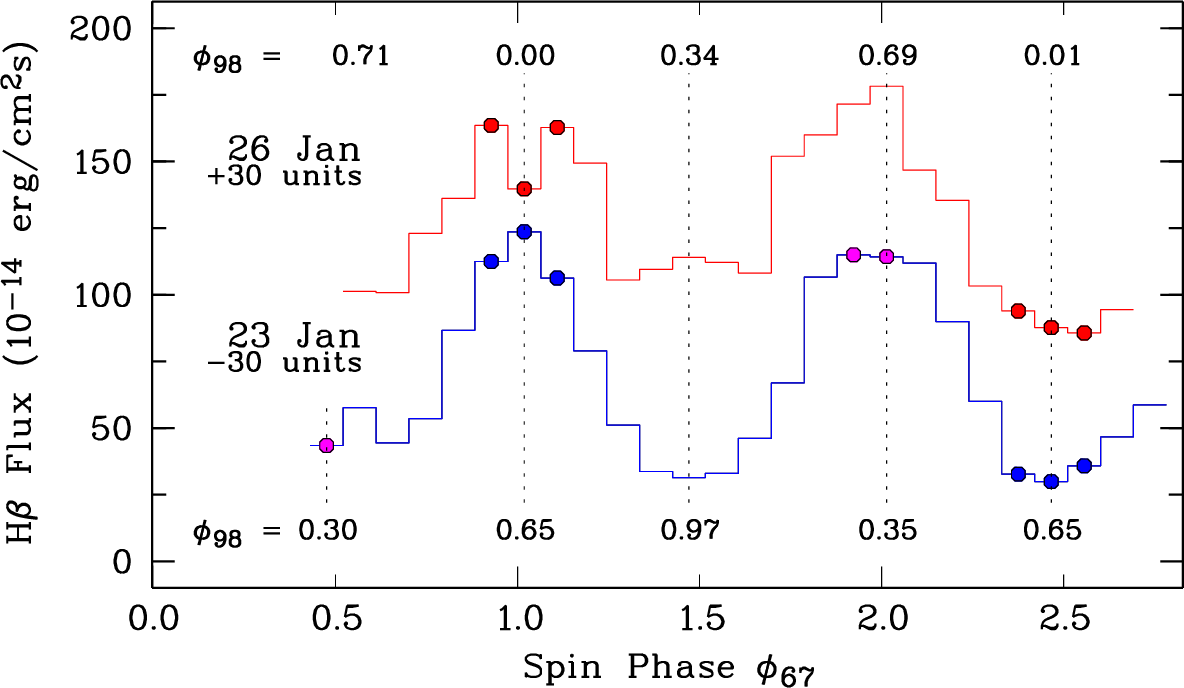}
\caption{
  Flux-calibrated spin light curves of the integrated \hbet\ line
  flux. The orbital phases are given in the figure. The colored dots
  indicate the same phase bins as in the left panel.}
\label{fig:fig3}
\end{figure}

The tomograms of \ion{Ca}{ii}$\lambda8498$ and
\ion{Fe}{i}$\lambda 4957$ in Fig.~\ref{fig:tomosec} demonstrate the
sole origin of the lines from the illuminated face of the secondary
star.  The weighted mean phase offset of the three metal lines is
$\langle\phi_0\rangle=0.0094(16)$ or
\mbox{$O\!-\!C=\!-0.00064(11)$}\,days (Table~\ref{tab:offset}). We
adopt the blue-to-red zero crossing as inferior conjunction of the
secondary star, which occurred at
\begin{equation}
    \phi_\mathrm{98} = \phi_{HS}+0.0094.
    \label{eq:phi98}
  \end{equation}  
This phasing is used to represent the orbital motion throughout this
paper.  
The light curve of January 26 in Fig.~\ref{fig:fig3} shows the
depression caused by the eclipse in time bin 6, coincident with the
first spin maximum. With a bin size of 5 min, we can only state that
the eclipse occurred within  0.02 in orbital phase from the
predicted times on the ephemerides of \citet{helliersproats92},
\citet{echevarriaetal16},\footnote{The ephemeris of
  \citet{echevarriaetal16} was approximately converted from HJD to
  BJD(TDB).} and Eq.~\ref{eq:phi98}.

\begin{table}[b]
  \caption{Observed times of events quoted as BJD(TDB) with errors,
    and time predicted by either the linear eclipse ephemeris of
    \citet{helliersproats92} or the quadratic ephemeris for the spin
    maxima of \citet[][their Eq. 1]{maucheetal09} }
\begin{tabular}{c@{\hspace{4mm}}c@{\hspace{10mm}}c@{\hspace{4mm}}c@{\hspace{4mm}}c} 
\hline \hline \noalign{\smallskip}
\multicolumn{2}{c}{Observed~~~~~~~~} &   \multicolumn{2}{c}{Reference~~~~~} & \\
 BJD (TDB)  &  Error  &  Cycle & BJD (TDB)  & $O-C$ \\
 $-2400000$ &   (d)   &        & $-2400000$ &  (d)  \\[1.0ex]
\noalign{\smallskip} \hline
\noalign{\medskip}
\multicolumn{5}{l}{\textit{Mean blue-to-red zero crossing} ~~~ \textit{Hellier \& Sproats (1992)}}\\[0.5ex]
53030.78990 & 0.00011 & 224681 & 53030.79054 & $-0.00064$ \\[1.0ex]
\multicolumn{2}{l}{\textit{Eclipse timing}}  & \multicolumn{3}{l}{\textit{~Hellier \& Sproats (1992)}} \\[0.5ex]
53030.79092 & 0.00150 & 224681 & 53030.79054 & $+0.00038$ \\[1.0ex]
\multicolumn{2}{l}{\textit{Spin maxima}} & \multicolumn{3}{l}{\textit{~Mauche et al. (2009)}}\\[0.5ex]
53027.76471 & 0.00100 & 329304 & 53027.76376 & $+0.00095$ \\
53027.81132 & 0.00100 & 329305 & 53027.81031 & $+0.00101$ \\
53030.79023 & 0.00150 & 328369 & 53030.78926 & $+0.00097$ \\
53030.83661 & 0.00100 & 329370 & 53030.83580 & $+0.00081$ \\[1.0ex]
\noalign{\smallskip} \hline      
\end{tabular}
\label{tab:offset}
\end{table}

The times of four spin maxima of the \hbet\ flux are listed in
Table~\ref{tab:offset} along with the times predicted by the quadratic
ephemeris of \citet{maucheetal09} and the corresponding \mbox{$O-C$}
values.
The mean $O-C$ is $0.00094(35)$
days or $-0.020(8)$ in $\phi_\mathrm{67}$. We adopt spin  phases
\begin{equation}
    \phi_\mathrm{67} = \phi_{M09}-0.020
\end{equation}
to match our observations.

\subsection{Flux calibration} 

The observations were performed in good seeing, although somewhat variable, which allowed an absolute calibration of the spectra. To this end,
we applied a seeing correction, based on the simultaneous information
provided by the Paranal Difference Image Motion
Monitor \citep[DIMM,][]{sarazinroddier90} and by the width of the individual
spectral images perpendicular to the dispersion direction. The two
methods yielded similar estimates of the  FWHM seeing that varied
between 0.9\arcsec\ and 2.2\arcsec\ on January 23 and between
0.7\arcsec\ and 1.2\arcsec\ on January 26.  We corrected each spectrum
for the seeing losses assuming a Gaussian point spread function and a
source centered in the 1.0\arcsec\ slit. We estimated the corrected
fluxes to have an internal accuracy of better than 20\%. Spectral
fluxes are quoted in units of $10^{-16}$\,\ergsa.

\section{Results}
\label{sec:spectra}

\subsection{Analysis of the phase-resolved spectra}
\label{sec:overview}

\begin{figure*}[t]
\includegraphics[height=54.0mm,angle=0,clip]{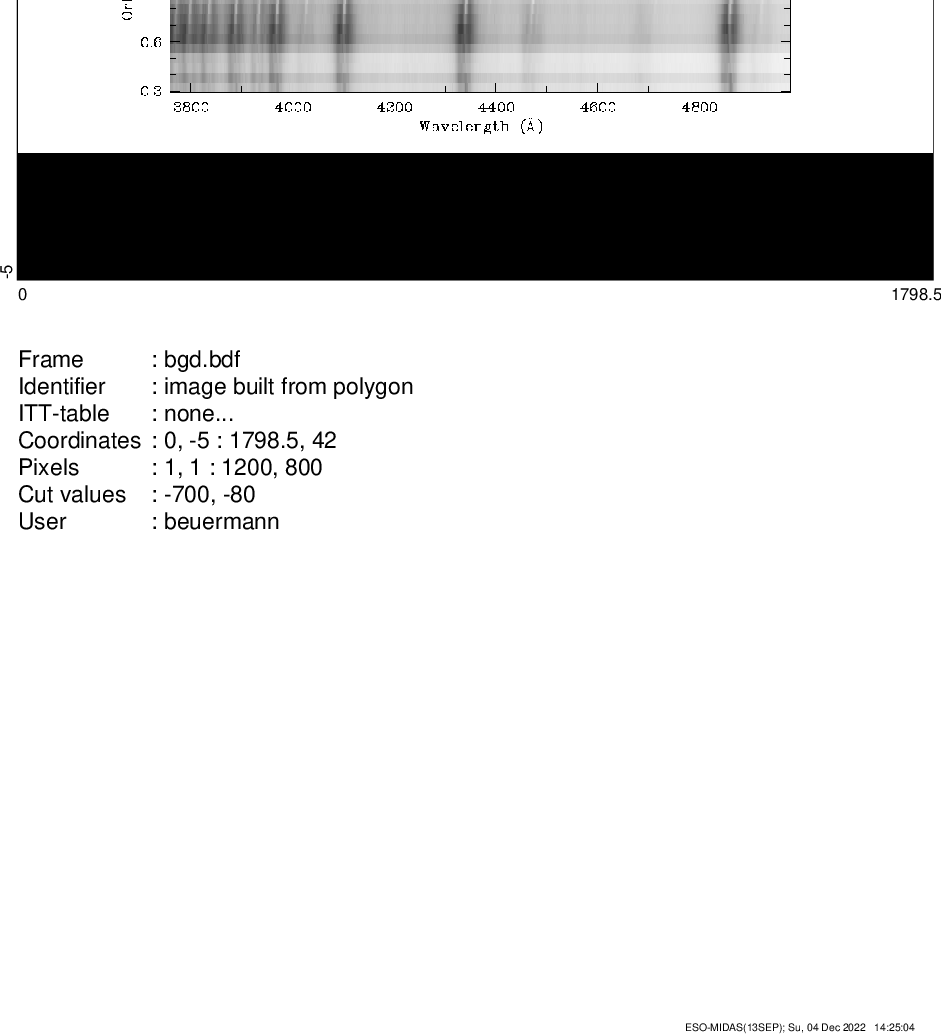}
\hfill
\raisebox{53.0mm}
{\includegraphics[width=53.0mm,angle=270,clip]{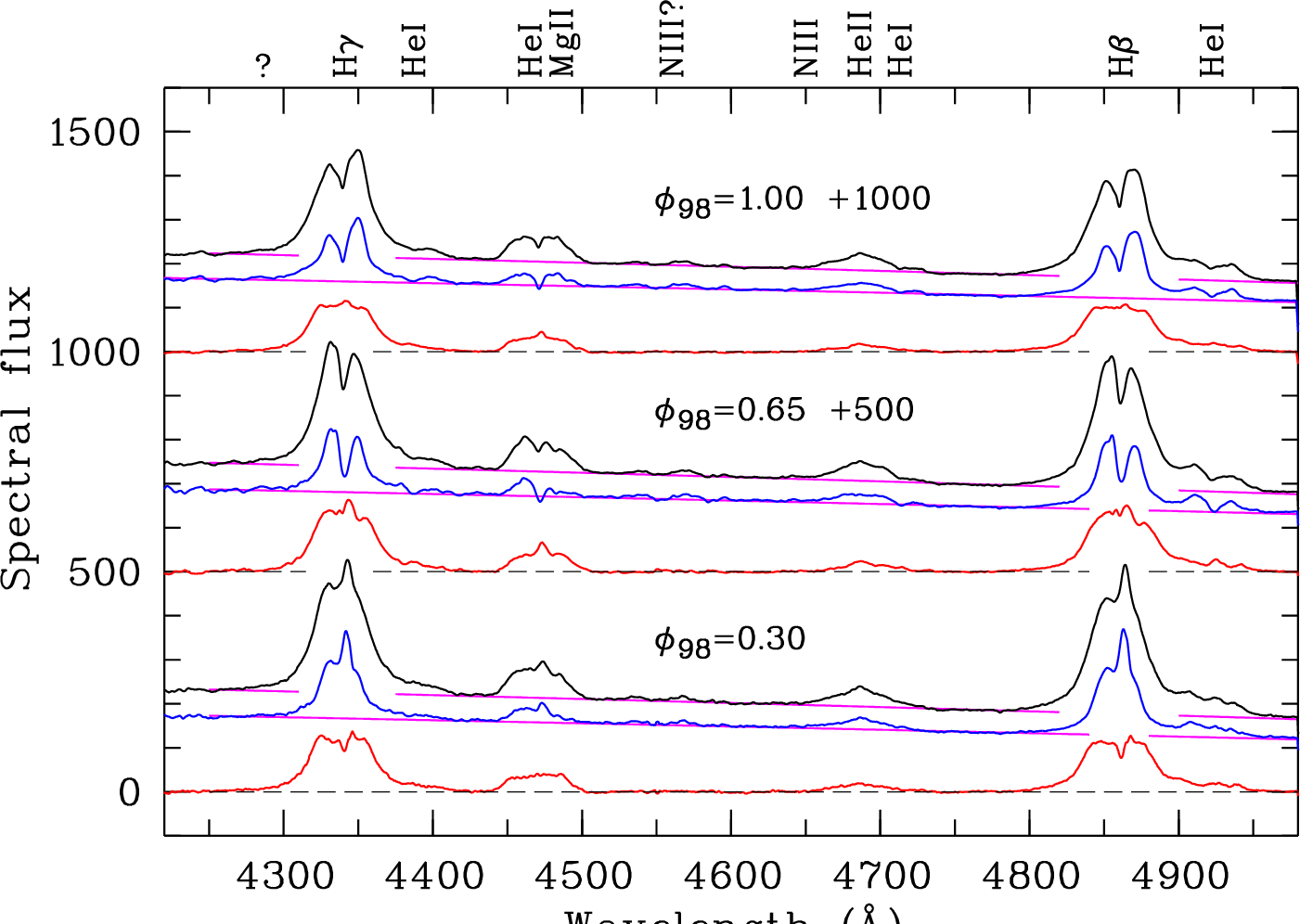}}
\caption{
  Flux-calibrated blue spectra taken on 23 and 26 January 2004.  Left:
  Grayscale representation of the first night, with spectral flux
  increasing from white to black. Right: Mean spectra of the \hgam\ to
  \hbet\ spectral region at spin maximum (black) and spin minimum
  (blue) for the orbital phases noted in the figure. The best-fit
  continua that represent the lower envelope to the spectra are shown
  as magenta curves. The spin modulation of the emission-line
  component is shown in red.}
  \label{fig:gray1}
\end{figure*}

Representative information on our spectral data is provided in
Figs.~\ref{fig:phase}, \ref{fig:fig3}, and \ref{fig:gray1}. The left
panel of Fig.~\ref{fig:gray1} shows the set of 26 flux-calibrated
spectra of 23 January 2004, which cover the Balmer lines \hbet\ to
H11.  On each of the two observing nights, 1.5 orbital periods and 2.2
spin periods were covered. Because of the near $3:2$ commensurability
of the two periods, coverage in the \phiorb$-$\phispin\ phase space is
sparse (left panel in Fig.~\ref{fig:phase}).  We decomposed these
trailed spectra into the spin modulated line wings and the orbital
phase-dependent line core. To this end, we proceeded in a two-step
process. In the first step, we constructed a preliminary spin template
as the difference between the spectral fluxes at spin maximum and spin
minimum, separated by 1.0 orbital and 1.5 spin periods. This choice
ensures the best possible subtraction of the S-wave component which
approximately repeats on the orbital period. Our data contain three
such instances, at \phiorb\,=\,0.30, 0.65, and 1.00. In the second
step, we noted that the line wings of the three preliminary spin
templates agree so closely that forming a mean template is
warranted. This is then employed to decompose the individual rows of
the trailed spectra of both observing nights.

\begin{figure*}[t]
\includegraphics[width=42.5mm,angle=270,clip]{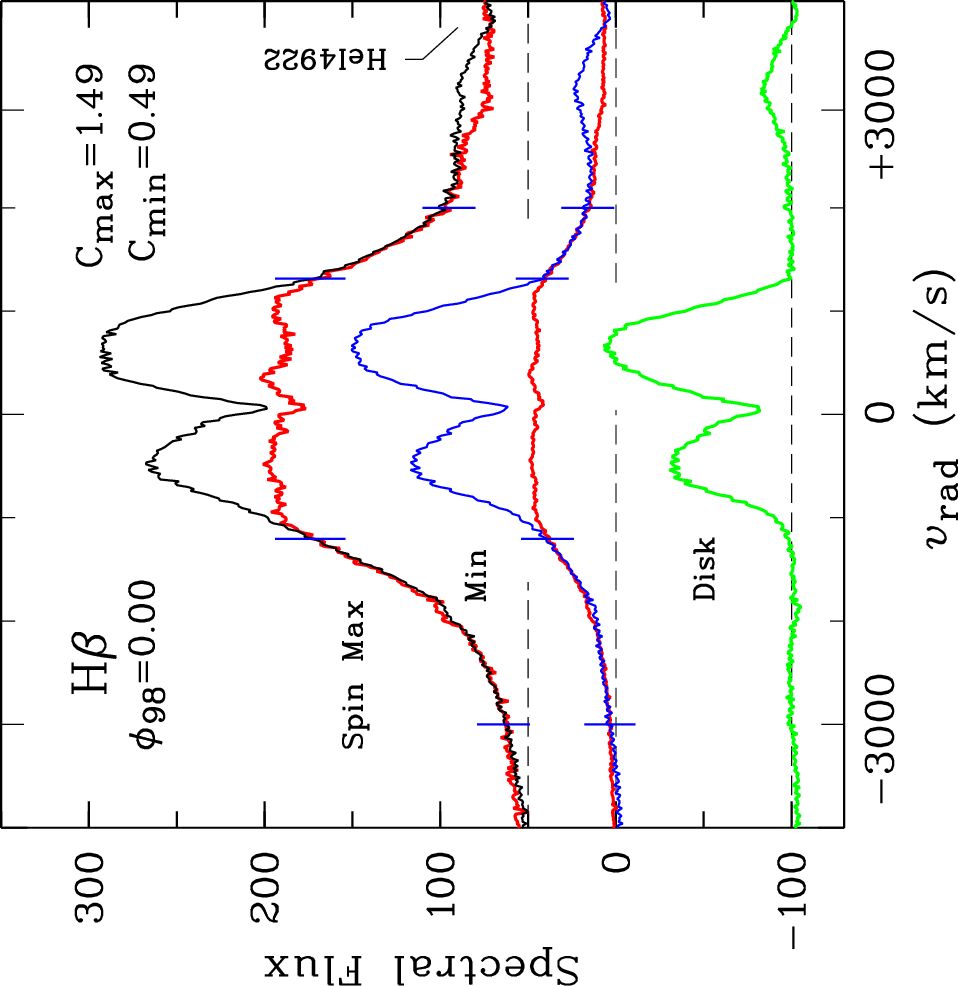}
\hspace{1mm}
\includegraphics[width=42.5mm,angle=270,clip]{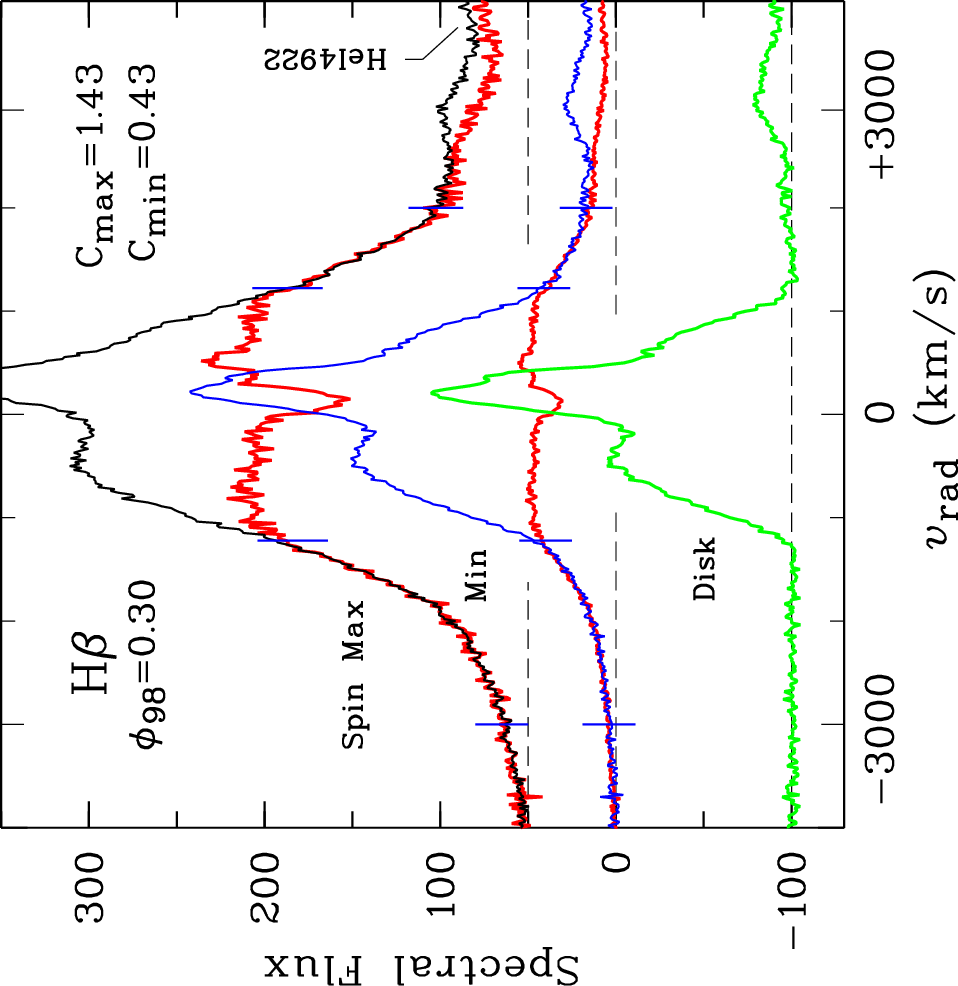}
\hspace{1mm}
\includegraphics[width=42.5mm,angle=270,clip]{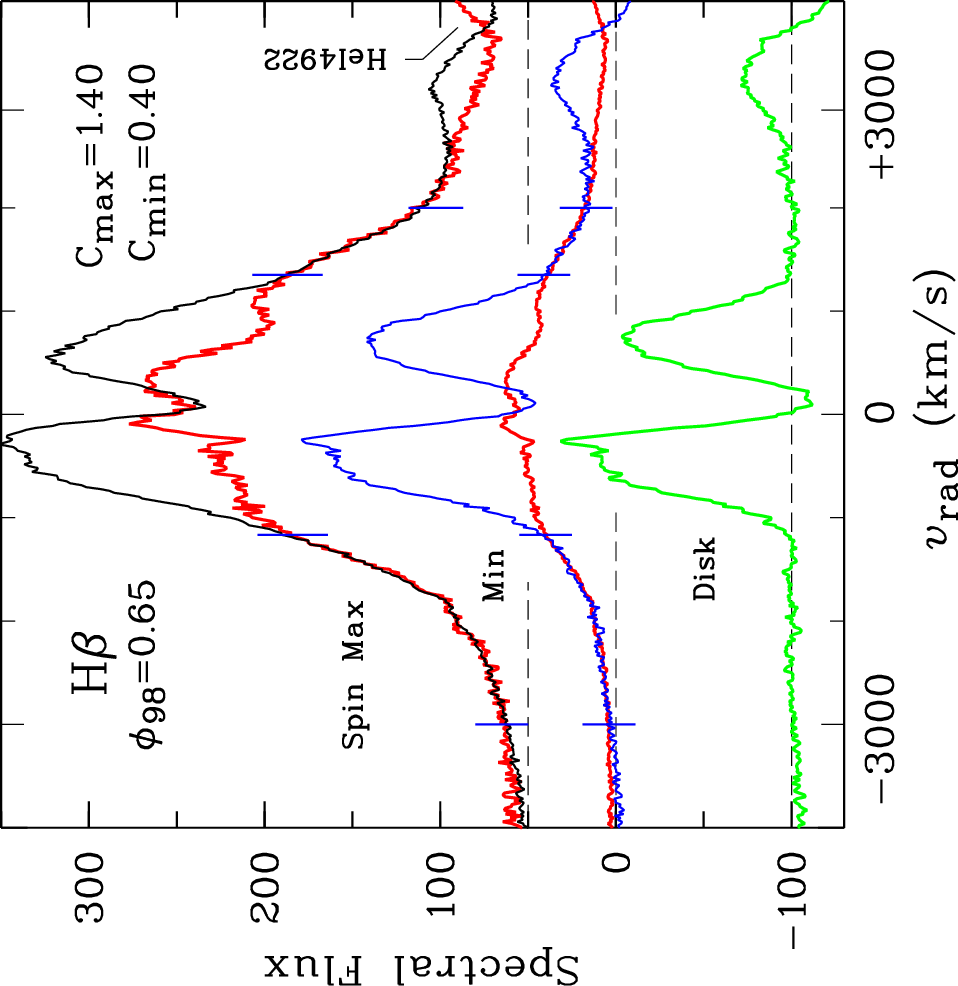}
\hfill
\includegraphics[width=42.5mm,angle=270,clip]{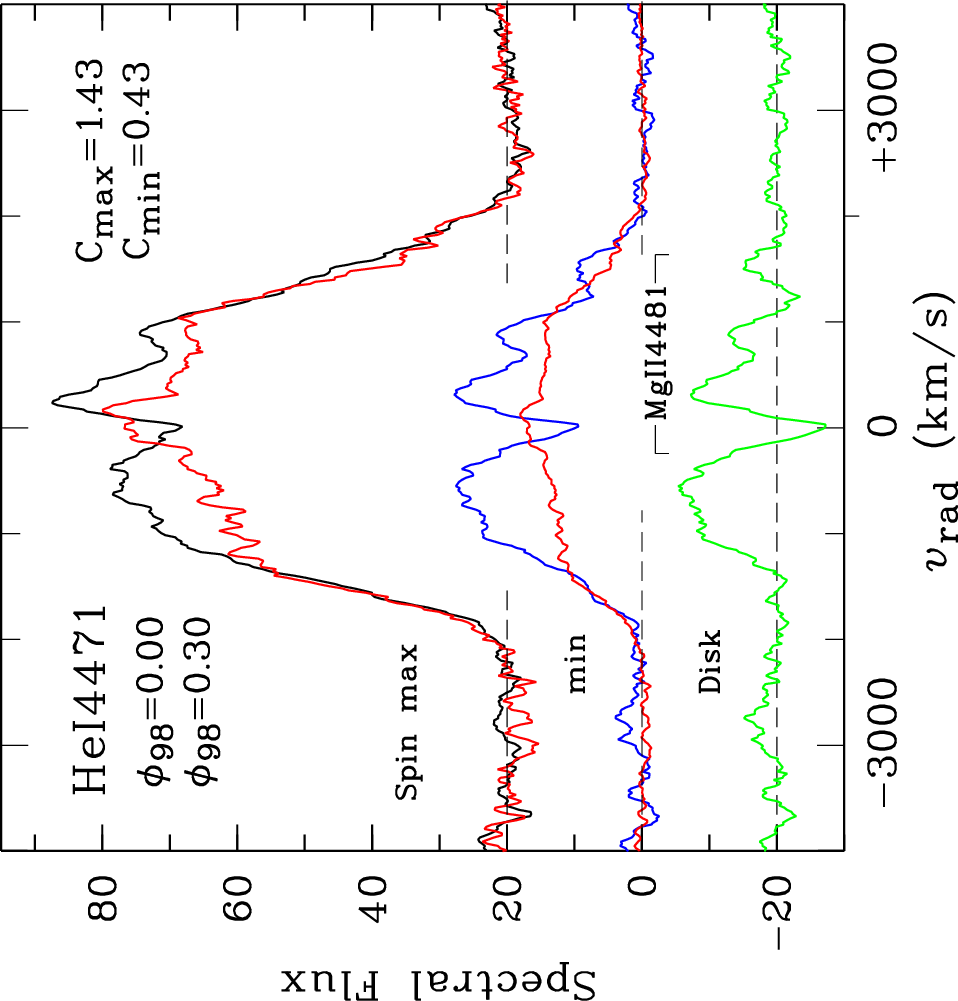}

\vspace{2mm}
\includegraphics[width=42.5mm,angle=270,clip]{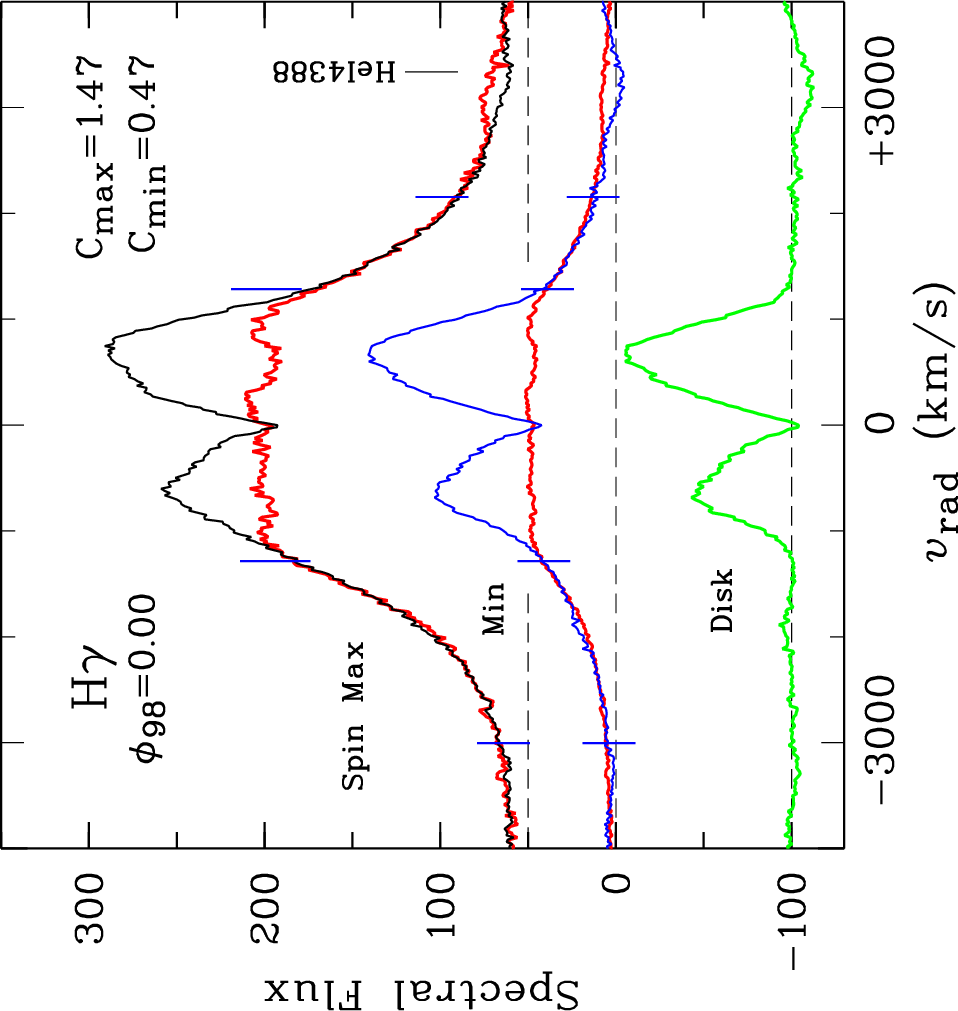}
\hspace{1mm}
\includegraphics[width=42.5mm,angle=270,clip]{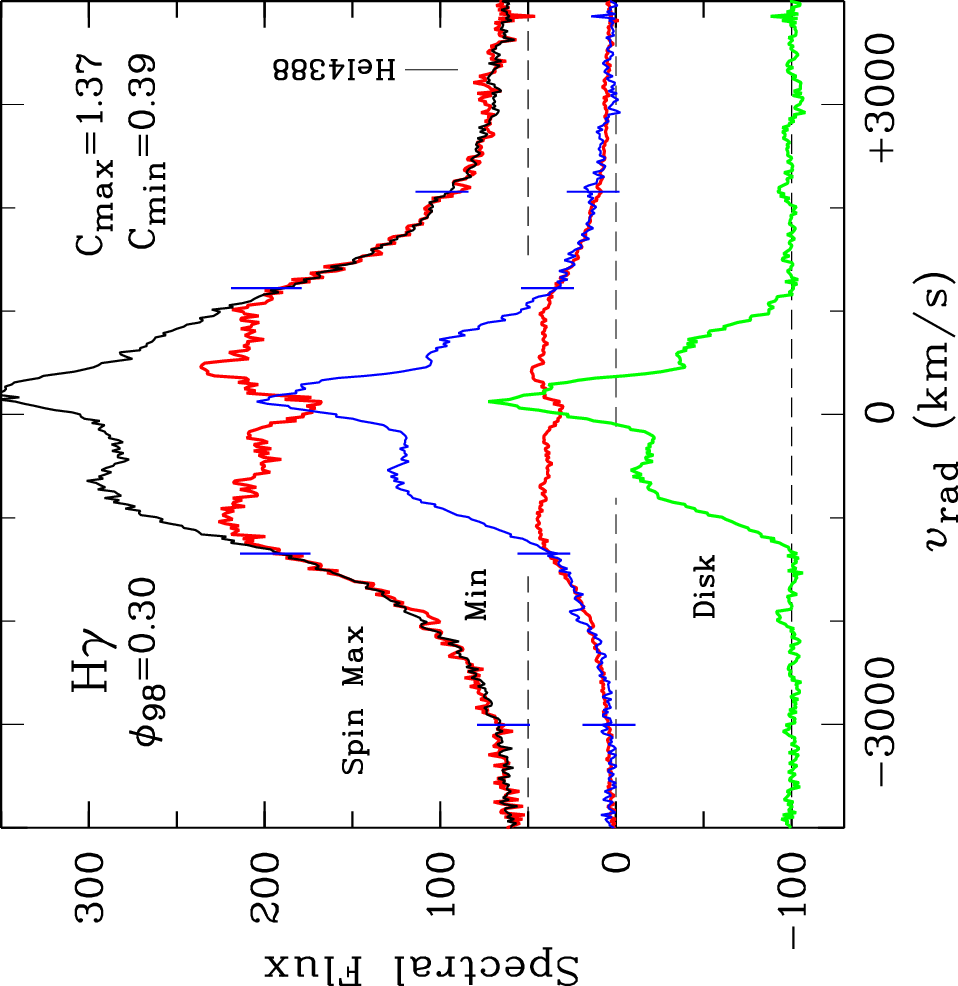}
\hspace{1mm}
\includegraphics[width=42.5mm,angle=270,clip]{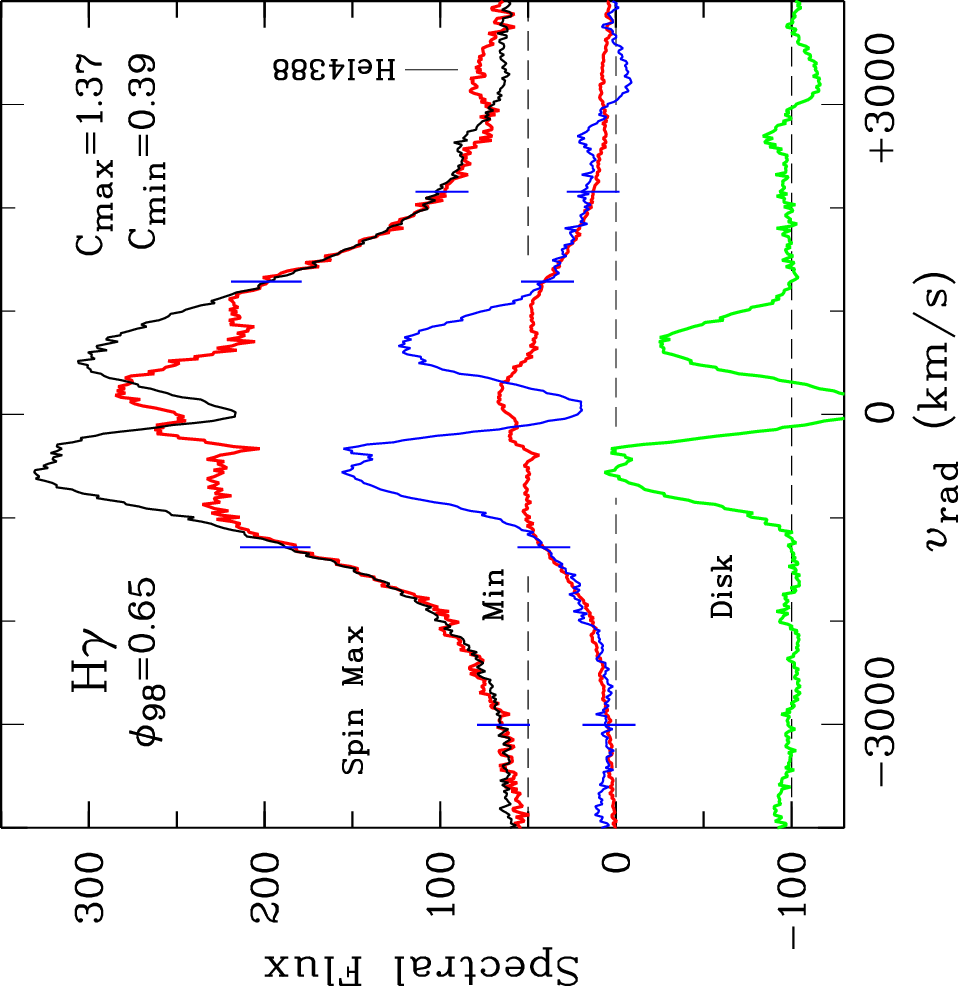}
\hfill
\includegraphics[width=42.5mm,angle=270,clip]{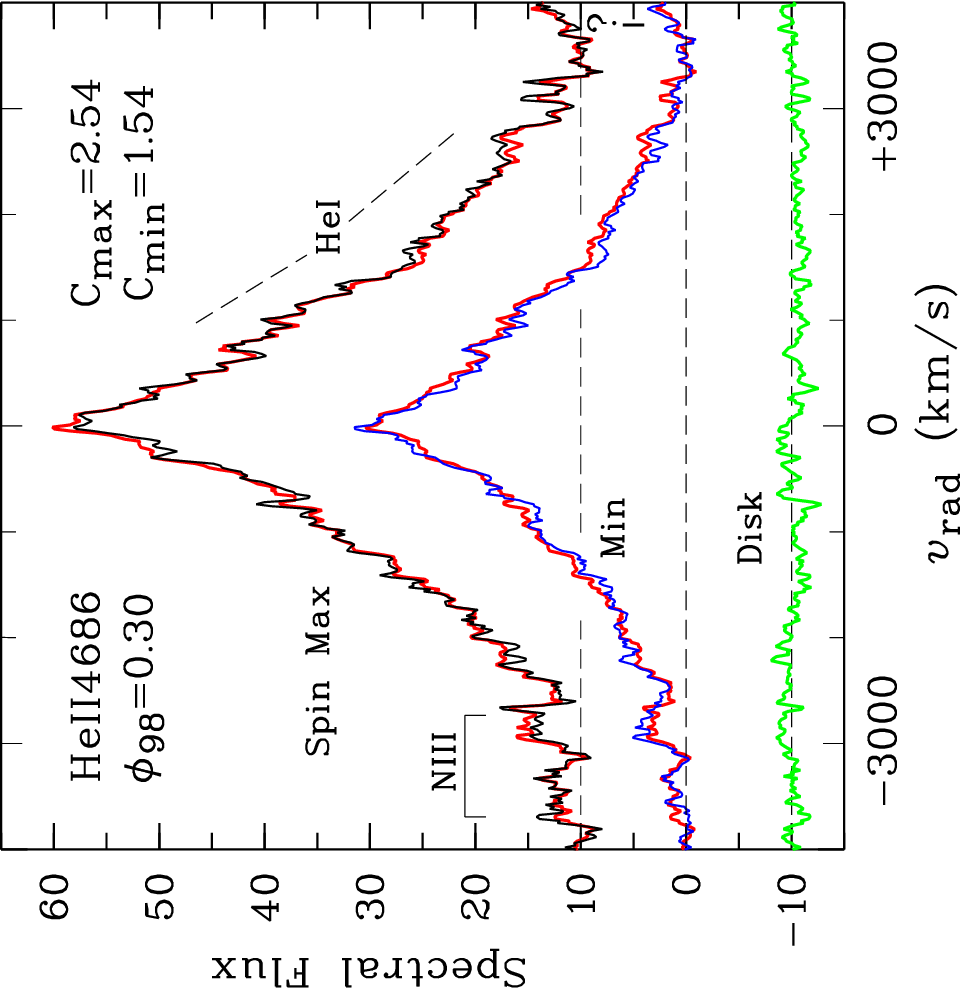}
\caption{\hbet\ and \hgam\ spin maximum
  and minimum spectra (black and blue, respectively) at
  \phiorb$\,=\!0.30, 0.65,$ and 1.0 (six left-most panels). Overplotted are the fitted
  spin-modulated components $f_\mathrm{mod}(\lambda )$ from Eq.~(4)
  with amplitudes $C_\mathrm{max}$ and $C_\mathrm{min}$ (red). Below,
  the derived disk component is shown (green). The two right panels show
  the decomposition of the \ion{He}{i}$\lambda 4471$ line at
  \phiorb$\,=\!0.0$ and 0.30 and of \ion{He}{ii}$\lambda 4686$ at
  \phiorb$\,=\!0.30$. No \ion{He}{ii} emission from the disk is seen,
  but S-wave emission is present at other orbital phases (see text).}
\label{fig:decom}
\end{figure*}

The three orbital phases selected for the first step of the template
construction are indicated by the intersections of the dashed lines in
the left panel of Fig.~\ref{fig:phase}. To improve the statistics, we
formed averages of up to three spectra around the respective
phases. These are marked in Fig.~\ref{fig:phase} by the magenta, blue,
and red dots, respectively. The resulting mean spectra at spin maximum
and minimum are displayed in the right panel of Fig.~\ref{fig:gray1}
as the black and blue curves, respectively.
Obtaining the spin templates requires subtraction of the underlying
continua from the observed spectra. The high quality of the UVES
spectra allows this process to be reliably performed longward of
4100\,\AA.  It is affected in the short-wavelength wing of \hdel\ by
weak disturbing lines and becomes increasingly difficult at shorter
wavelengths because of the merging line wings (see
Fig.~\ref{fig:gray1}, left panel). Consequently, we restricted the
analysis to \hbet\ and \hgam. The fitted continua are depicted as the
magenta curves in the right panel of Fig.~\ref{fig:gray1}. The spin
modulation, which is  the difference between the emission line components
at spin maximum and minimum, is included in the  form of the red
curves. Larger representations of the \hbet\ and \hgam\ line profiles
on a velocity scale are shown in Fig.~\ref{fig:decom}.

\subsection{Decomposing the  Balmer lines at  \phiorb=0.3, 0.65, and 1.0}
\label{sec:decom}

We find that the entire line wings vary to a high degree of accuracy
by a spin-phase dependent numerical factor only (i.e., they are
multiples of the spin modulation),
\begin{equation}
  f_\mathrm{mod}(\lambda )=f_\mathrm{obs,max}(\lambda )-f_\mathrm{obs,min}(\lambda ),
  \label{eq:decom1}
\end{equation}
which we adopt as a preliminary template at the selected orbital
phase (red curves in Figs.~\ref{fig:gray1}). We express the 
line profiles as
\begin{eqnarray}
  f_\mathrm{obs,max}(\lambda ) & = & C_\mathrm{max}f_\mathrm{mod}(\lambda +\delta\lambda_\mathrm{max}) +\delta f_\mathrm{max} + f_\mathrm{core}(\lambda ), \nonumber\\
  f_\mathrm{obs,min}(\lambda ) & = & C_\mathrm{min}f_\mathrm{mod}(\lambda + \delta\lambda_\mathrm{min}) + \delta f_\mathrm{min} + f_\mathrm{core}(\lambda ),
\label{eq:decom2}
\end{eqnarray}
\noindent and perform the fit only over the line wings. Here
$\delta\lambda$ allows for a small shift of the template in
radial-velocity and $\delta f$ for an error in subtracting the
continuum or a deviation from the assumed proportionality of the line
wings to $f_\mathrm{mod}$.  For vanishing $\delta\lambda$ and
$\delta f$, the wings would vary strictly proportionally to
$f_\mathrm{mod}$, and the fit parameters $C_\mathrm{max}$ and
$C_\mathrm{min}$ would be related by
\begin{eqnarray}
  C_\mathrm{min}=C_\mathrm{max}-1.
  \label{eq:decom3}
\end{eqnarray}
\noindent The validity of our model depends on the degree to which the
parameters $C_\mathrm{max}$ and $C_\mathrm{min}$ obtained from the fit
comply with Eq.~\ref{eq:decom3}. The last term in Eq.~\ref{eq:decom2}
designates the core component that is assumed to be the same at spin
maximum and minimum and to contribute negligibly in the wings. It is
obtained by subtracting the fitted $Cf_\mathrm{mod}$ and $\delta f$
from the observed line profile.

The model fits for \hbet\ and \hgam\ at the three selected orbital
phases are displayed in the six left panels of
Fig.~\ref{fig:decom}, now with radial velocity as the abscissa. The
black and blue curves are the continuum-subtracted line profiles at
spin maximum and minimum. The red curves represent the adjusted
spin modulation fitted to both wings over the wavelength ranges
indicated by the vertical tick marks. At negative velocities the fits
extend to $-3000$\,\kms\ and at positive velocities to about
2100\,\kms, avoiding the disturbing HeI lines. The fit excludes the
central part of the line, where the fitted $f_\mathrm{mod}$ deviates
drastically from the observed profile. Appropriate limits are
$\pm\,(1250\!\pm\!50)$\,\kms\ from the line center.  The fits are
excellent. The fit parameters $C_\mathrm{max}$ and $C_\mathrm{min}$
(noted in the figures) follow Eq.~\ref{eq:decom3} faithfully. The
absolute terms $\delta f$ are close to zero with an average
$\langle \delta f \rangle\!=\!0.5\!\pm\!1.0$ in ordinate units. The
wavelength shifts $\delta\lambda$ stay below $\pm1.2$\AA, equivalent
to $\mid\!\upsilon_\mathrm{rad}\!\mid\,<80$\,\kms, consistent with
orbital motion and a moderate spin-dependent radial-velocity
modulation (see Sect.~\ref{sec:system}). Given that \oex\ is an IP,
the highest velocities of the spin component certainly result from the
magnetically guided accretion stream near the WD, and since we found
that the line wings  represent a single entity, this conclusion
necessarily holds for the entire spin-modulated component. That the
line wings should vary only by a spin-phase dependent factor is not
self-evident, but can be understood if spin modulation of the line
flux arises mainly from occultation of part of the emission. The core
or disk component is clearly limited to radial velocities between
$\pm\,1300$\,\kms. It is displayed at the bottom of the respective
figure as the green curve.

The wings of the templates for the three orbital phases agree so
closely that it is warranted to combine them into a common final
template applicable to all orbital phases. This allows us to decompose
the entire body of the trailed spectra and to subject the derived
phase-dependent core component to a tomographic analysis. From the
tomograms presented in Sect.~\ref{sec:tomo}, we obtain further insight
into the structure of the inner accretion disk.
 
\subsection{Final templates for the Balmer line spin components}
\label{sec:decom1}

We show the normalized mean of the three preliminary templates of
\hbet\ and \hgam\ in Fig.~\ref{fig:temp} (blue curves). The central
region between $\pm1200$\,\kms\ is roughly flat-topped, possibly the
result of radiative transfer of the Balmer lines in puffed-up matter
of the transition region between the  disk and magnetosphere and the bulge
of the disk. The remaining fluctuations at the top of the profiles can
probably be assigned to imperfect repetition of the S-waves in
consecutive orbits. We therefore approximate the central part of the two
templates by a flat top (black curves), and adopt these
curves as the final spin templates of \hbet\ and \hgam. In their
wings they contain remnants of the nearby \ion{He}{i} lines; of
additional unidentified faint broad lines of disk origin (Disk); and
of narrow metal lines, mostly $\ion{Fe}{i}$, from the secondary star
(Fe\,I). These are all contained in the trailed spectra as well. Only
the short-wavelength wing of \hbet\ is practically free of such
disturbances and optimally represents the extended Balmer line wings.

\subsection{Decomposing the trailed spectra of \hbet\ and \hgam}
\label{sec:decom2}

\begin{figure}[t]
\includegraphics[width=41.5mm,angle=270,clip]{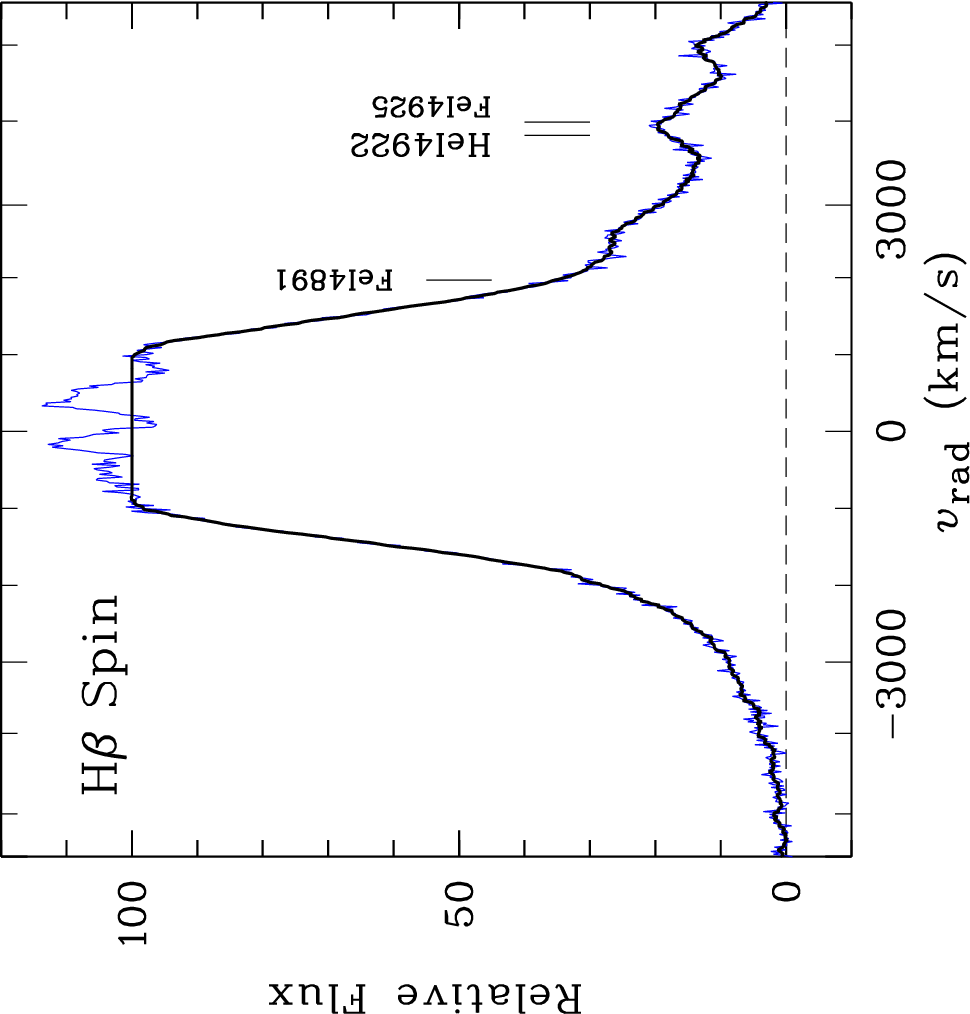}
\hspace{1mm}
\includegraphics[width=41.5mm,angle=270,clip]{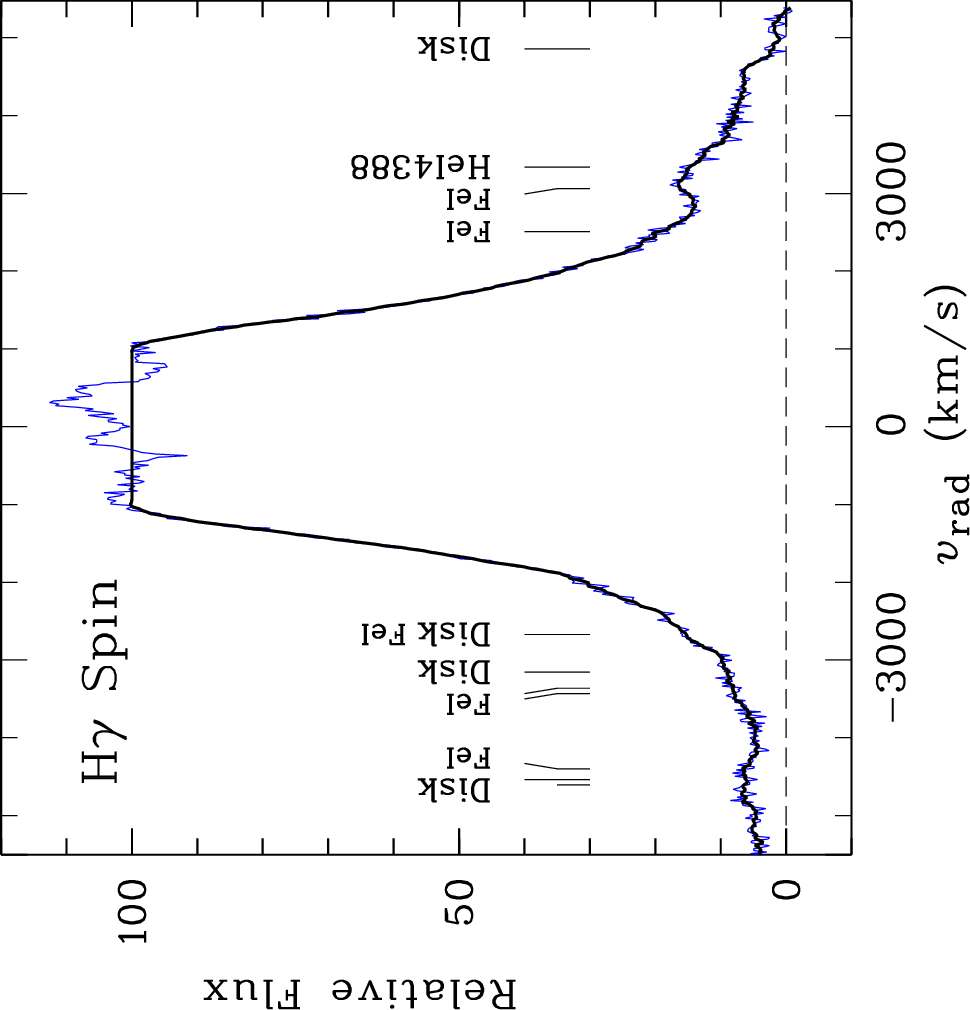}
\caption{Mean templates of the spin-modulated component of \hbet\ and
  \hgam.}
\label{fig:temp}
\end{figure}

\begin{figure*}[t]
\includegraphics[height=85mm,angle=0,clip]{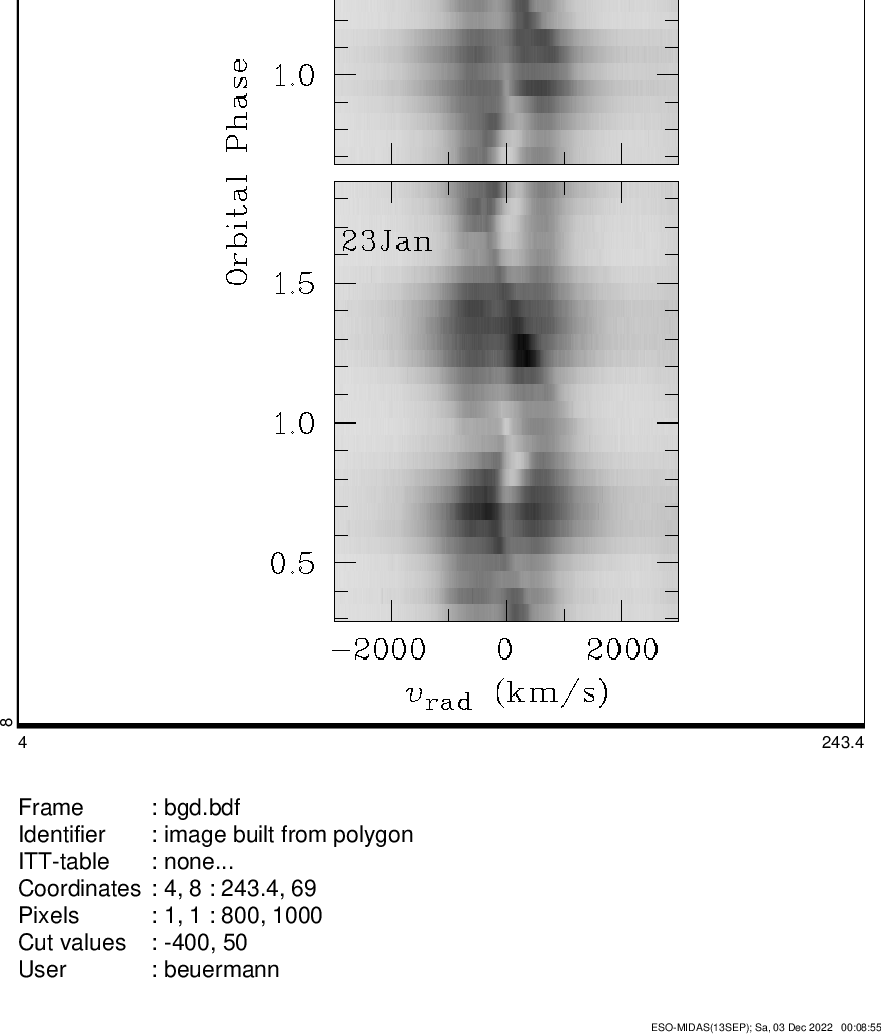}
\hspace{0.0mm}
\includegraphics[height=85mm,angle=0,clip]{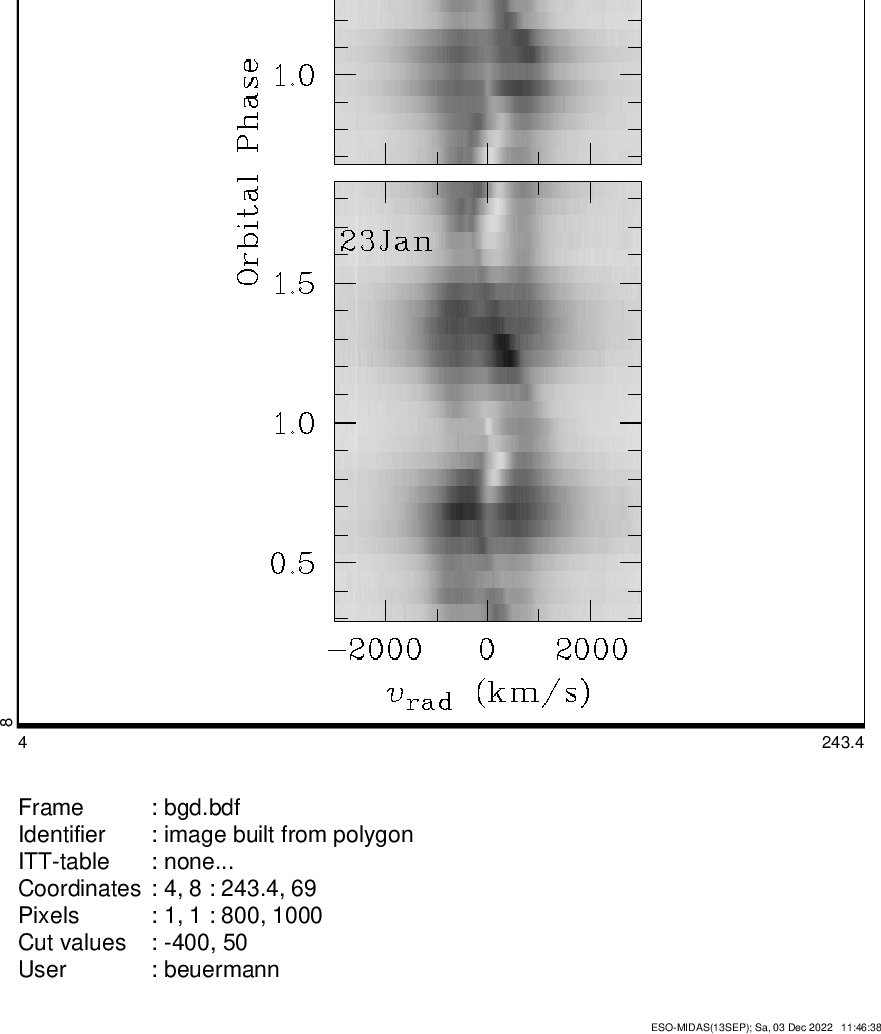}
\hspace{0.0mm}
\includegraphics[height=85mm,angle=0,clip]{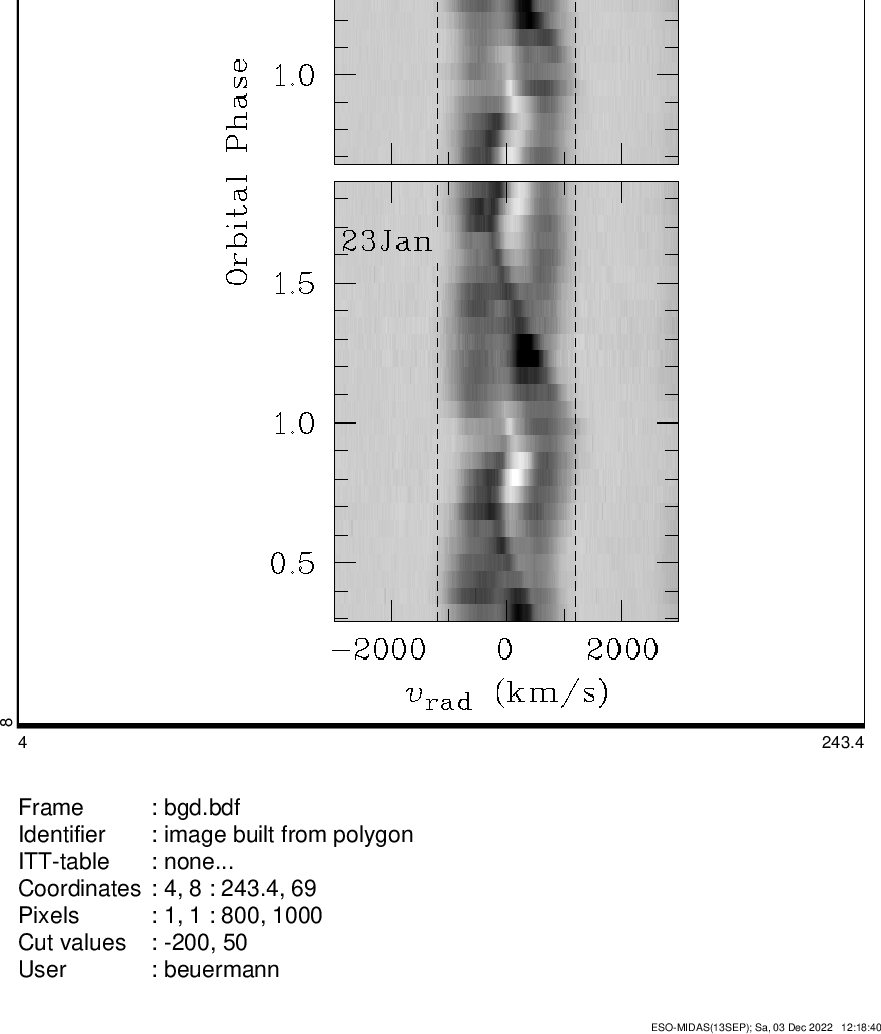}
\hspace{0.0mm}
\includegraphics[height=85mm,angle=0,clip]{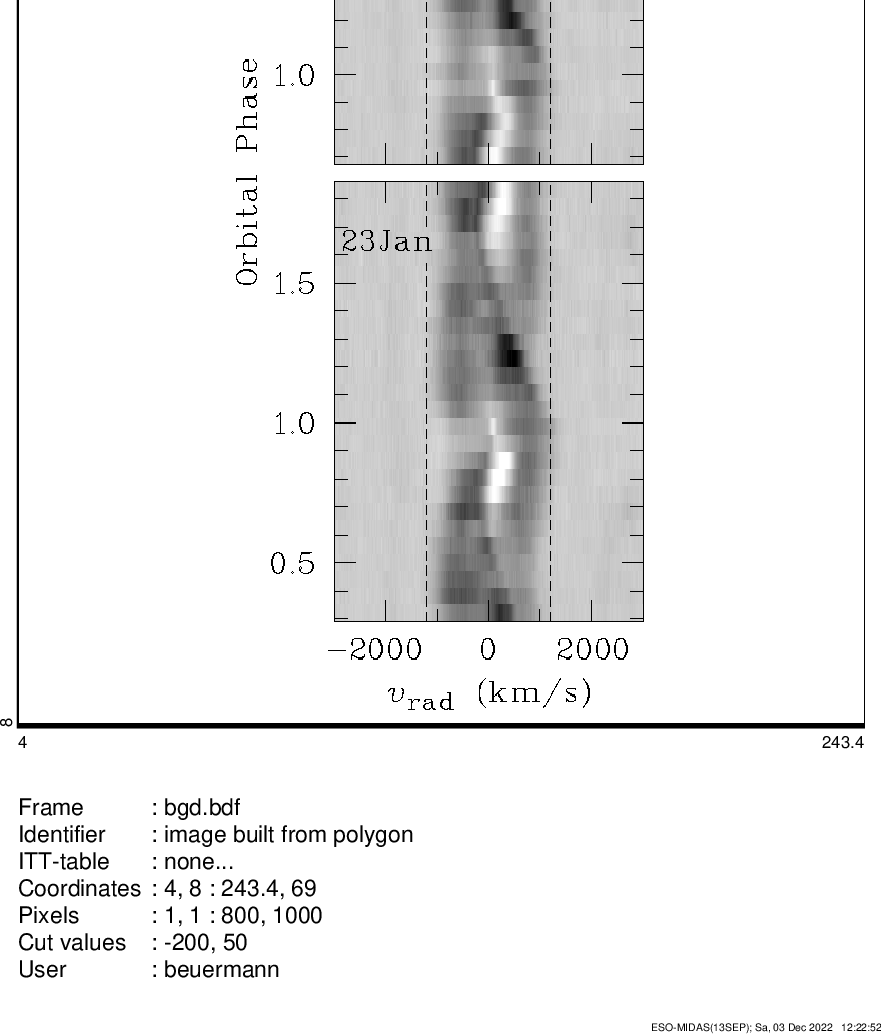}
\hspace{0.0mm}
\includegraphics[height=85mm,angle=0,clip]{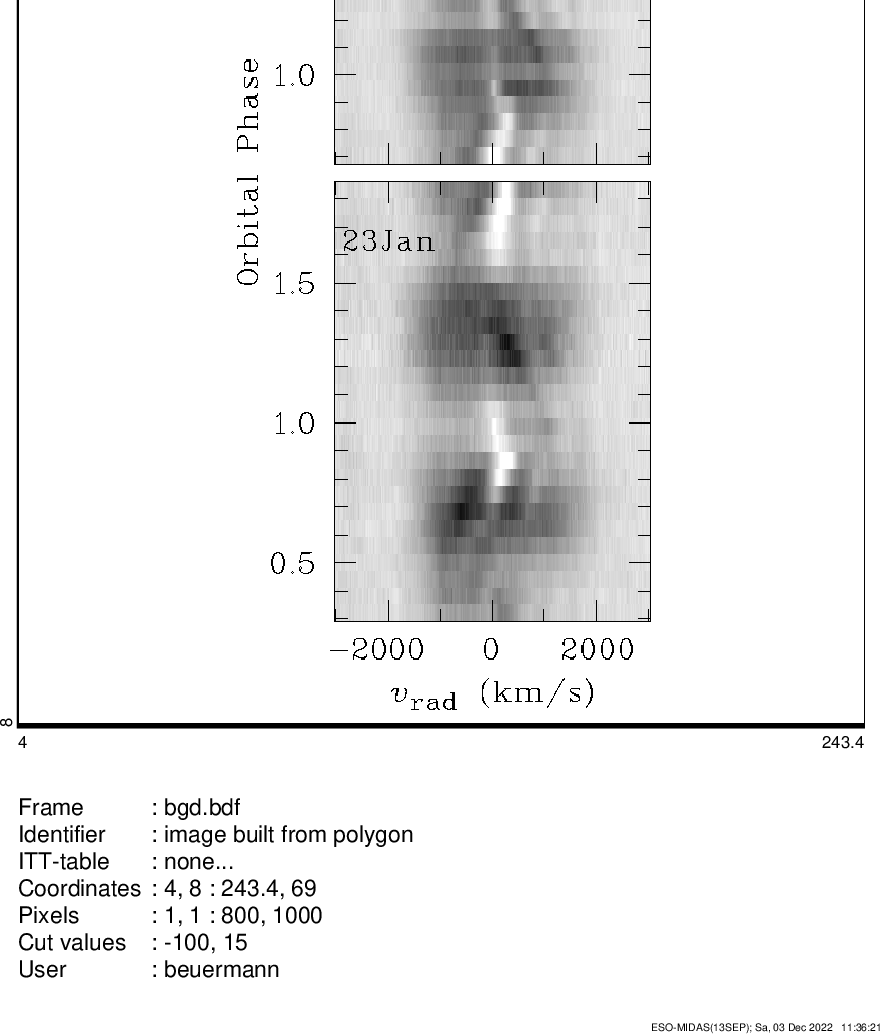}
\hspace{0.0mm}
\includegraphics[height=85mm,angle=0,clip]{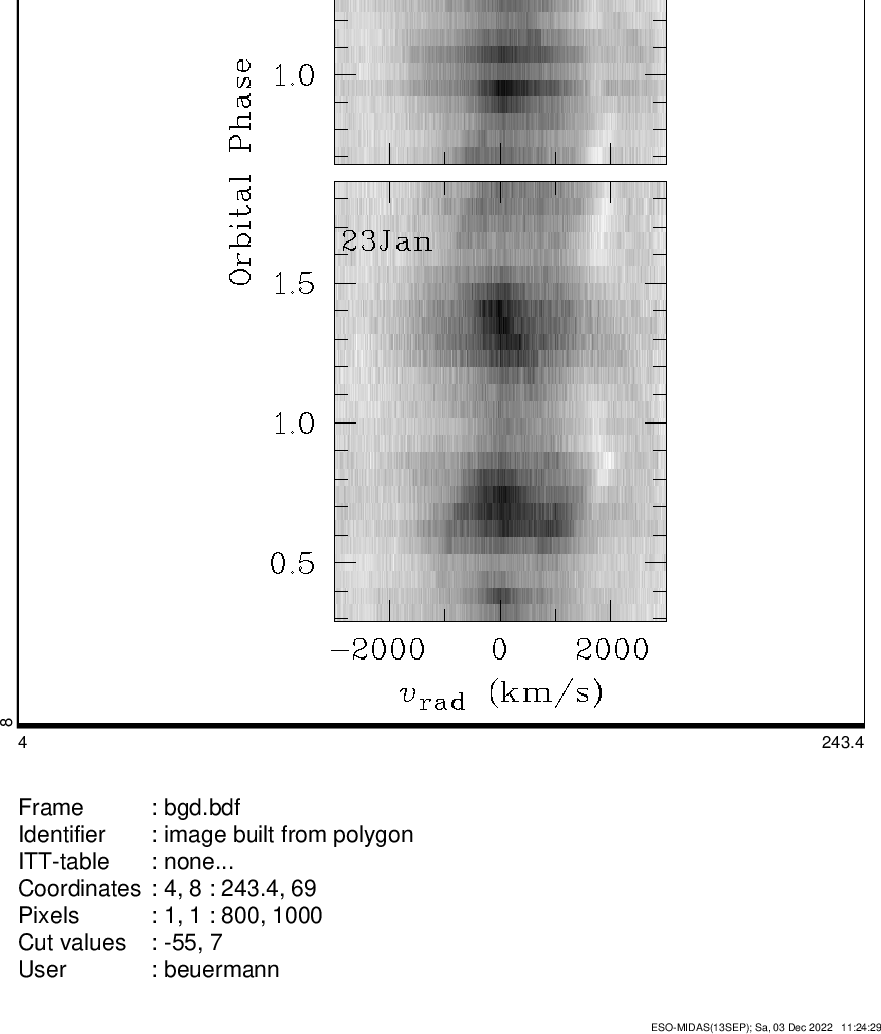}
\caption{
  Trailed hydrogen and helium emission line spectra.  Left two panels:
  Observed line profiles of \hbet\ and \hgam. Center two panels: Disk
  component obtained by subtracting the fitted spin component from the
  observed profiles (see text). The emission of this component stays
  between the two vertical dashed lines at $\upsilon_\mathrm{term}=\pm1200$\,\kms.
  Right two panels: Trailed line spectra of \ion{He}{i}\,$\lambda4471$
  and \ion{He}{ii\,$\lambda4686$}. These are disturbed by
  \ion{Mg}{ii}\,$\lambda4481$ and \ion{He}{i}$\,\lambda4713$ centered
  at $+650$\,\kms\ and $+1750$\,\kms, respectively.}
\label{fig:gray2}
\end{figure*}

Using the templates of Fig.~\ref{fig:temp} to represent
$f_\mathrm{mod}(\lambda )$ in Eq.~\ref{eq:decom2}, we decomposed the
trailed \hbet\ and \hgam\ spectra in the left two panels of
Fig.~\ref{fig:gray2}. The fits were extended over the same intervals
that were used in deriving the templates (tick marks in
Fig.~\ref{fig:decom}). The result after subtracting the fitted spin
component are labeled  in the center panels of Fig.~\ref{fig:gray2} as
``\hbet\ disk'' and ``\hgam\ disk.'' The dashed lines indicate the
$\pm1200$\,\kms\ limit, outside of which the disk component rapidly
vanishes. The available evidence shows that this limit also holds   for
the higher Balmer lines (not shown). Since the profile of the template
is smooth, all statistical fluctuations of the original spectra are
assigned to the disk component. The fitted spin component is
featureless and consists of identical rows that vary in intensity with
spin phase and display small wavelength shifts. It contains no
information of relevance for a tomographic study \citep{hellier99} and
is not shown.  The radial velocities of the individual rows vary with
orbital and spin phase and are discussed in more detail in
Sect.~\ref{sec:system}.

\subsection{Decomposition of HeI$\lambda4471$ and  HeII$\lambda 4686$}
\label{sec:heii}

The two right panels in Fig.~\ref{fig:gray2} show the complex
continuum-subtracted trailed spectra of \ion{He}{i}$\lambda 4471$ and
\ion{He}{ii}$\lambda 4686$. Both are heavily disturbed by competing
lines, \ion{He}{ii} by \ion{He}{i}$\lambda 4713$ at a separation of
+27\,\AA\ and \ion{He}{i}$\lambda 4471$ by a previously unrecognized
contamination by \ion{Mg}{ii}$\lambda4481$ at $+10$\,\AA.

Despite the heavy contamination, applying the model of
Eq.~\ref{eq:decom2} to the \ion{He}{i}$\lambda 4471$ spectra at
orbital phases \phiorb$\,=\!0.0$ and 0.30 (Fig.~\ref{fig:gray2})
provides some important information. The spin-modulated component
extends only to wavelengths of $\pm2100$\,\kms\ (upper right panel of
Fig.~\ref{fig:decom}), much less than found for the Balmer lines or
\ion{He}{ii}$\lambda 4686$. Since the extreme wings of all lines
originate primarily from photoionization deep in the magnetospheric
accretion curtain or funnel, the difference between the helium lines
is likely the result of the ionization structure in the funnel: helium
exists almost entirely as \ion{He}{ii} close to the WD and
increasingly as \ion{He}{i} farther out. The central low-velocity
part of the \hei\ lines originates in part from collisional ionization
and photoionization in the disk, leading to a  double-peak
structure similar to that  in the Balmer lines. This does not hold for
\ion{He}{ii}$\lambda 4686$, however.
  
As has long been known, the line profile of \ion{He}{ii}$\lambda 4686$
differs from the Balmer and \ion{He}{i} lines, being single peaked,
with a base width of at least 6000\,\kms, comparable to that of the
Balmer lines. The \ion{He}{ii} line lacks the double peaks of disk
origin and the low-velocity absorption between \phiorb$\,=\!0.6$ and
0.9 that is prominent in the Balmer and \hei\ lines (white patches in
Fig.~\ref{fig:gray2}). Such absorption patches are seen in the
\ion{He}{ii} trailed spectrum at $+1750$\,\kms; they are created by
\ion{He}{i}$\lambda 4713$. Nevertheless, there is a weak component of
disk origin also in \ion{He}{ii}$\lambda 4686$ in the form of S-wave
emission from the bulge on the disk at \phiorb$\simeq\!0.65$
(Fig.~\ref{fig:gray2}).  At \phiorb$\,\simeq\!0.30$ an almost
undisturbed spin component of the \ion{He}{ii} line is observed, which
can be analyzed with Eq.~\ref{eq:decom2}. The lower right panel of
Fig.~\ref{fig:decom} shows the spectrum at spin maximum (black) and
spin minimum (blue) together with the multiples of the spin
modulation, fitted over $\pm3000$\,\kms\ (red). With
$C_\mathrm{max}=2.54$ and $C_\mathrm{min}=1.54$, \heii\ has a smaller
relative spin amplitude than the Balmer lines. In this case, the
entire line profiles at spin maximum and minimum are proportional to
each other. Hence, by Eq.~\ref{eq:decom2} there is no discernible disk
component at \phiorb$\,=\!0.30$ (green). The Balmer and \heii\ spin
profiles agree in the wings, but differ drastically between
$\pm\!1200$\,\kms. We conjecture that differences in the radiative
transfer are responsible. Pushed up matter at the inner edge of the
disk may prevent soft X-rays from reaching the surface of the thin disk, but
not the pushed-up matter at the outer edge of the disk where the
S-wave is produced. At an inclination of 78\degr, Balmer line emission
from the disk passes through denser parts of the bulge closer to the
orbital plane than \heii\ photons from the vicinity of the WD. Hence,
the \heii\ line may more closely correspond to the intrinsic spin
profile than the spin templates of Fig.~\ref{fig:temp}.
\citet{kimbeuermann96} showed that single-peaked line profiles
occurred in their model calculations, but the remaining systematic
uncertainties prevented definite conclusions.
  
\begin{figure*}[t]

\vspace*{1mm}
\includegraphics[height=68.0mm,angle=0,clip]{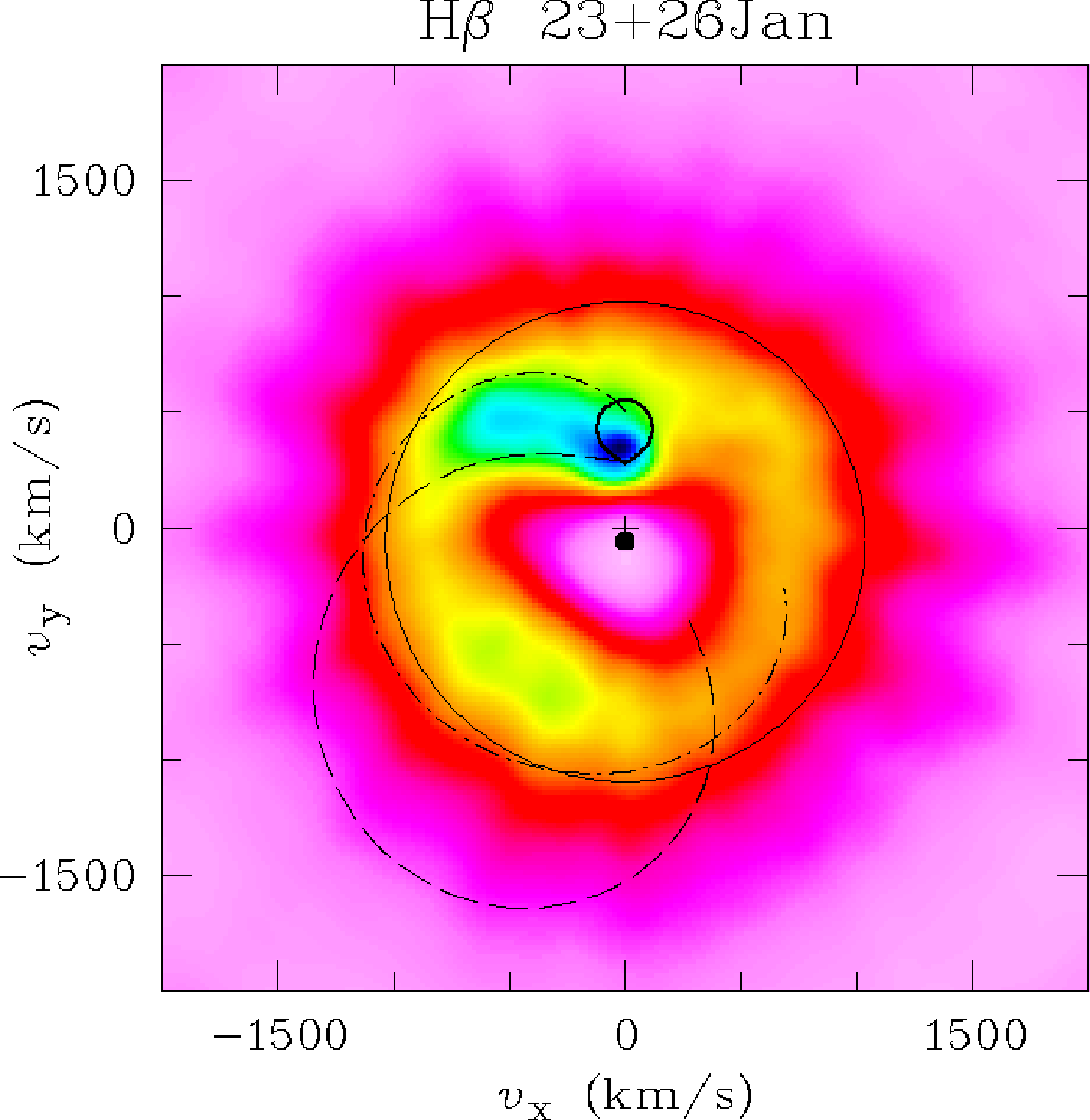}
\hspace{1.0mm}
\includegraphics[height=68.0mm,angle=0,clip]{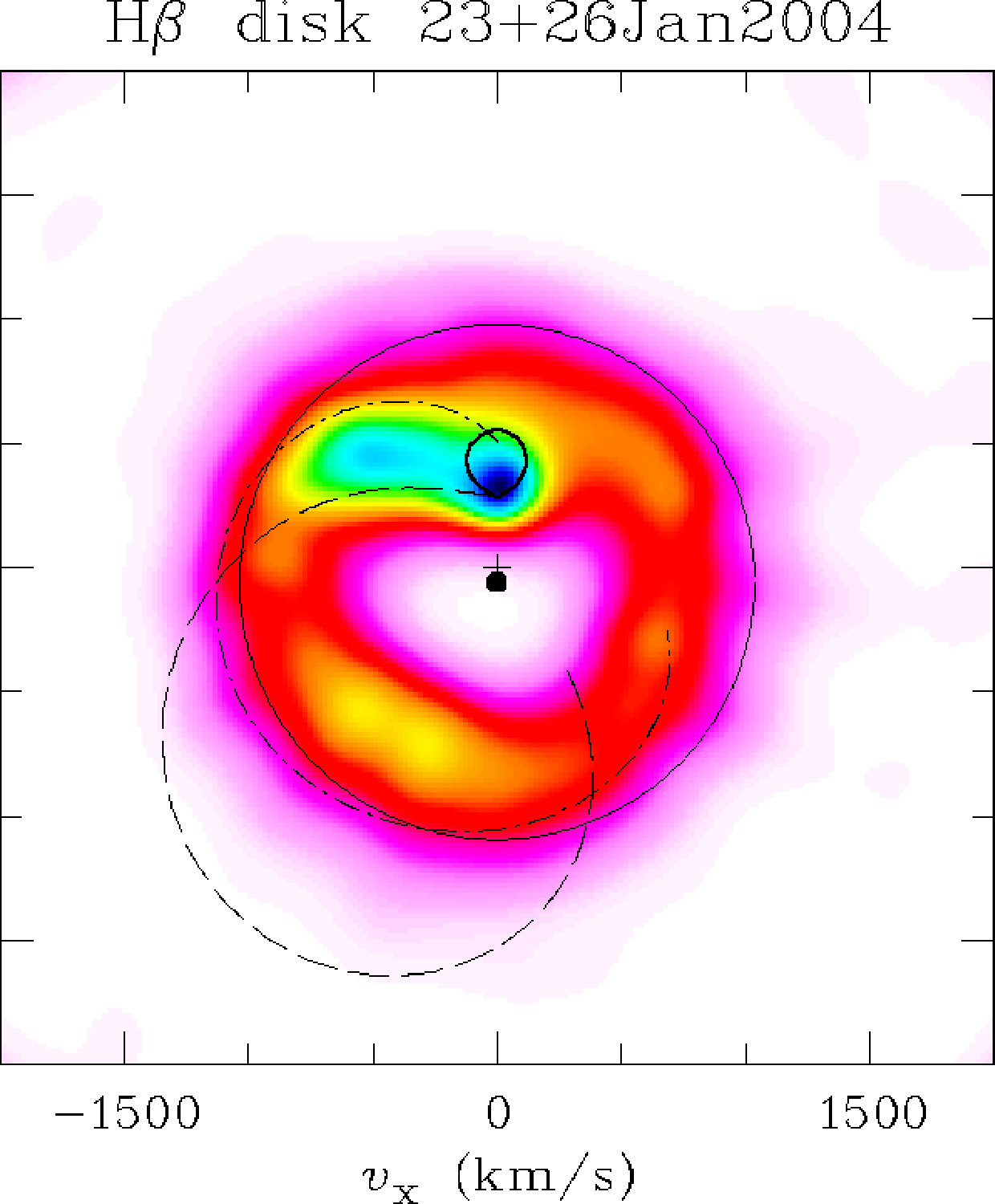}
\hfill
\includegraphics[height=68.0mm,angle=0,clip]{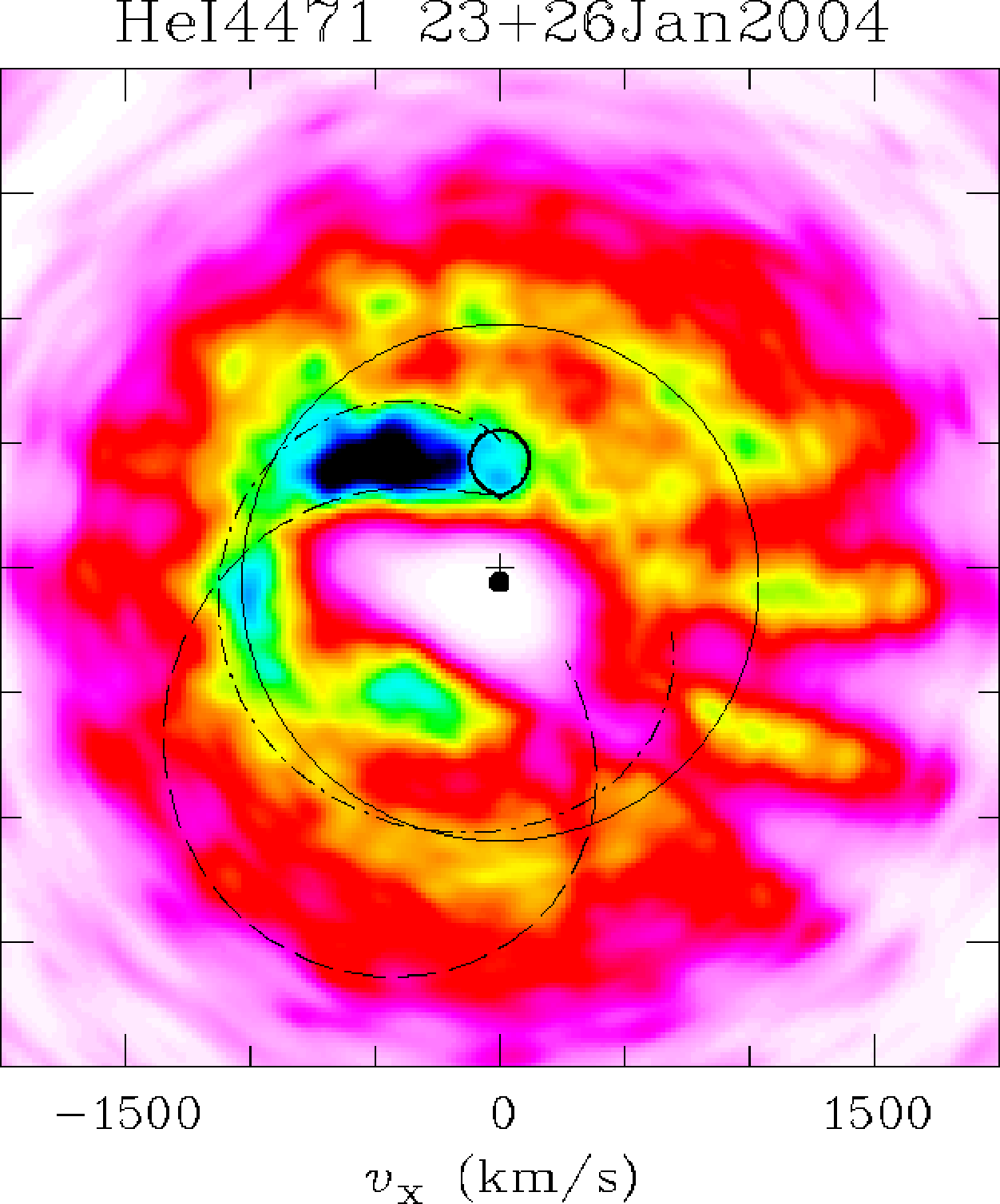}

\vspace*{2.0mm}
\includegraphics[height=69.2mm,angle=0,clip]{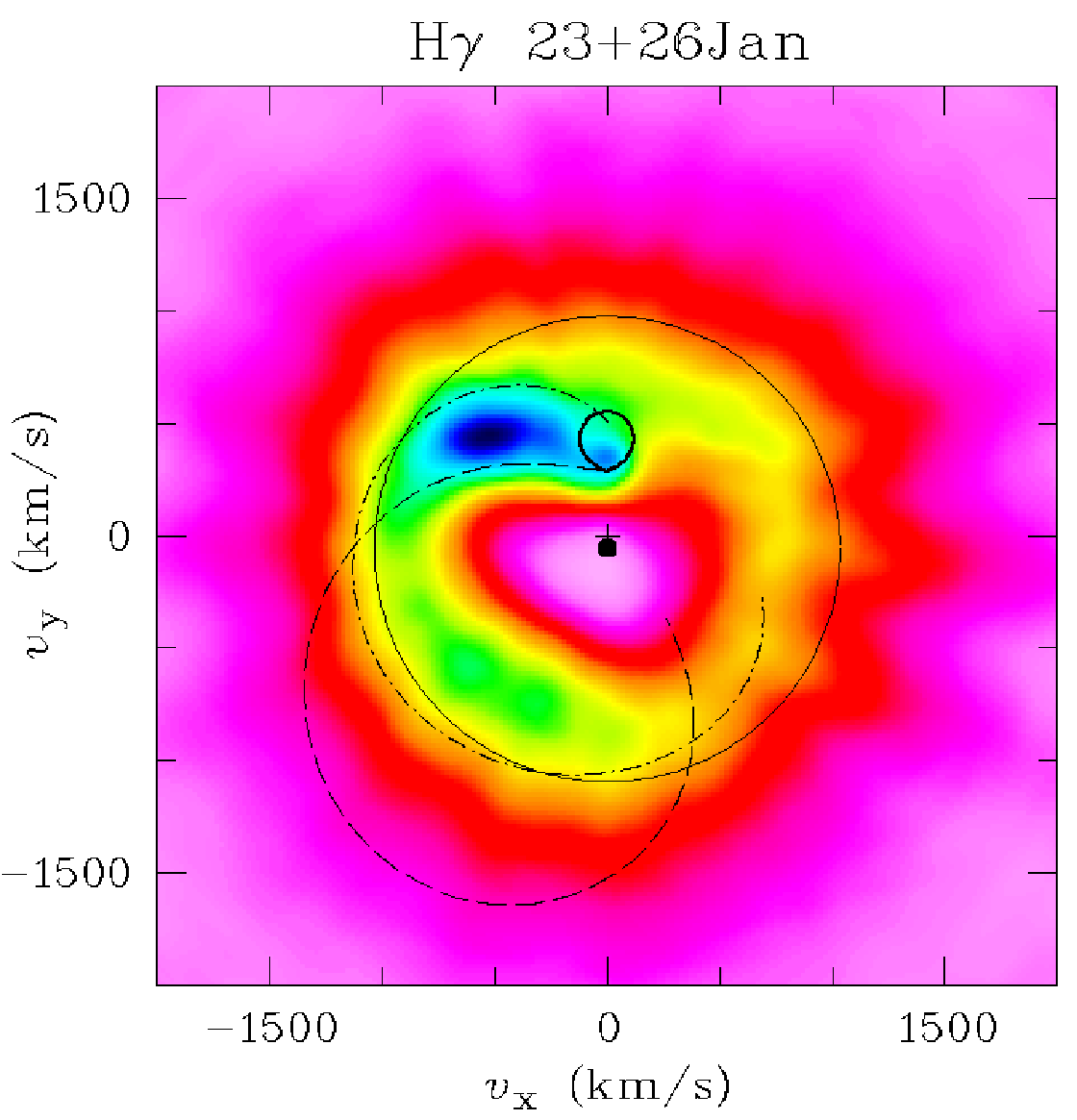}
\hspace{0.1mm}
\includegraphics[height=68.0mm,angle=0,clip]{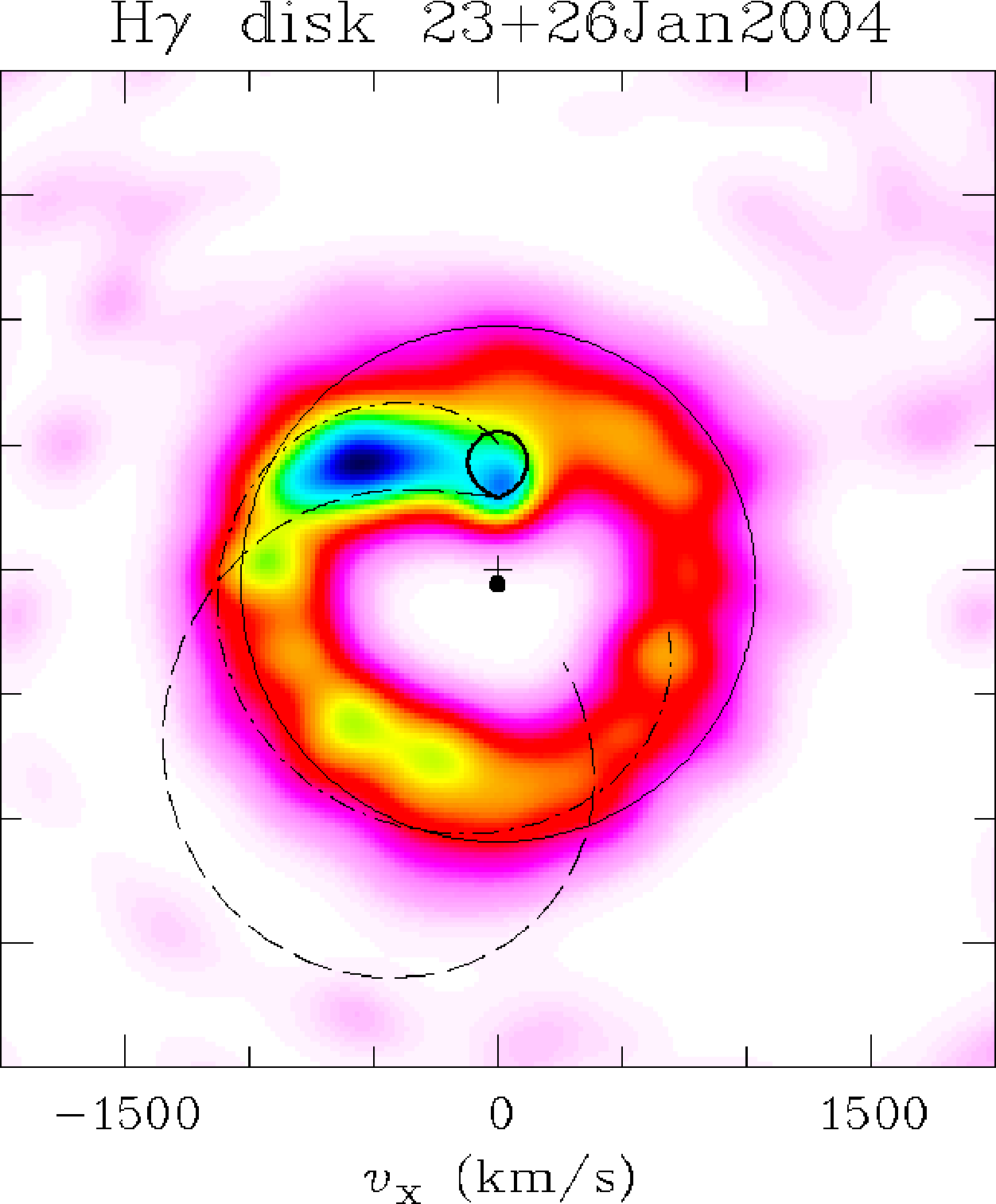}
\hfill
\includegraphics[height=67.6mm,angle=0,clip]{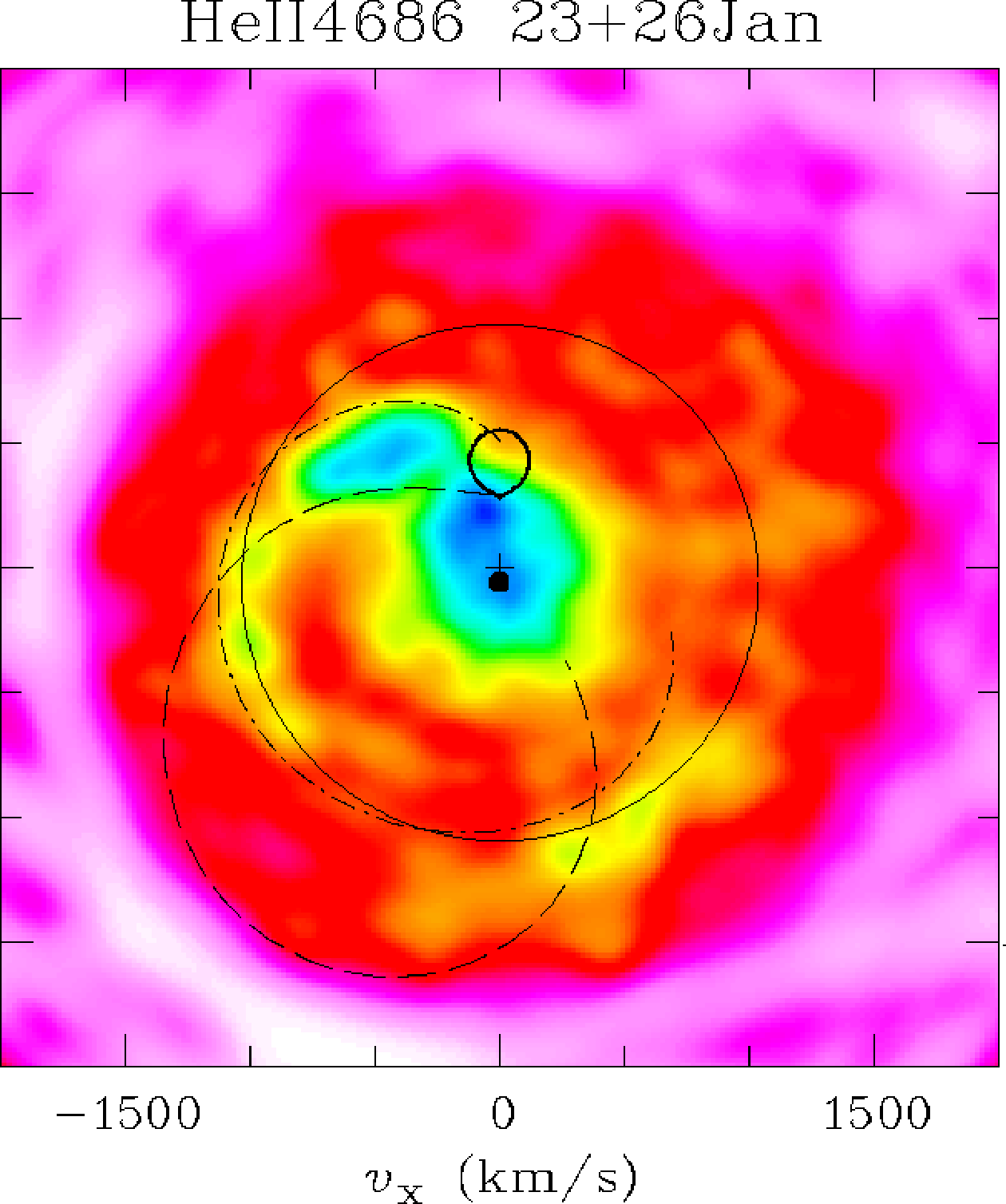}
\caption{Orbital tomograms of emission lines of \oex. Top, from
    left: (1) Tomogram of the observed \hbet\ spectra, (2) Tomogram
  of the disk component of \hbet\ after subtraction of the
  spin-modulated component, (3) Tomogram of the observed spectra of
  \ion{He}{i}\,$\lambda4471$. Bottom, from
    left: (4) Tomogram of the observed \hgam\ spectra, (5) Tomogram
  of the disk component of \hgam, (6) Tomogram of the observed spectra
  of \ion{He}{ii}\,$\lambda4686$. In each panel the dashed curve indicates the
  single-particle trajectory and the dot-dashed curve the Kepler
  velocity along that path.  The solid circle denotes the velocity at
  the circularization radius, shifted by $-59$\,\kms\ in
  $\upsilon_\mathrm{y}$ to be centered on the WD. The color scale is
  as follows: white=0.0, magenta$\simeq0.10$, red$\simeq0.25$,
  yellow$\simeq0.40$, green$\simeq0.55$, cyan$\simeq0.70$, light
  blue$\simeq0.80$, dark blue$\simeq0.90$, black$=1.0$.}
\label{fig:tomo}
\end{figure*}

\subsection{Doppler tomography}
\label{sec:tomo}


We subjected the combined trailed spectra of both observing nights in
Figs.~\ref{fig:tomosec} and \ref{fig:gray2} to a tomographic analysis
by fast maximum entropy Doppler imaging, which combines the advantages
 of the maximum entropy method with the speed of filtered
back-projection inversion \citep{spruit98}. The results are displayed
in Figs.~\ref{fig:tomosec} and \ref{fig:tomo}.  A bin size of
20\,\kms\ was chosen, which corresponds to 1.1 spectral bins of
0.3\,\AA\ in the observed spectra. The origin of the coordinate system
is at the center of gravity (plus sign), the WD is at
$\upsilon_\mathrm{y}=-59$\,\kms\ (bullet). The Roche lobe of the
secondary star is based on the system parameters of
Sect.~\ref{sec:system}, and is shown as the thick solid curve in all
tomograms.
The dashed and dot-dashed curves in Fig.~\ref{fig:tomo} describe the
single-particle trajectory in velocity space and the Kepler velocity
along that trajectory, respectively. The solid circle, which is
centered on the WD, denotes the line-of-sight component of the Kepler
velocity at the circularization radius. It is based on the fit of
\citet{hessmanhopp90} to the calculations of \citet{lubowshu75}, which
also account   for the gravitational pull of the secondary star,
\rcirc/$a=0.0859q^{-0.426}$, with $a$ the binary separation. For a
mass ratio $q=M_2/M_1=0.137$ (Sect.~\ref{sec:system}) the
circularization radius is \rcirc$=0.200\,a=\ten{9.44}{9}$\,cm. The
Kepler velocity at \rcirc\ is
$\upsilon_\mathrm{kep}(r_\mathrm{circ})=1055$\,\kms, with a line-of-sight component of 1032\,\kms\ for an inclination $i=77\fdg8$
(Sect.~\ref{sec:system}).

We used smoothing parameters that avoid spokes as much as possible,
without losing important structure. Spokes are prominent in the
tomogram of \hei$\lambda 4471$, but less so in the others.
Trailed spectra that vary at two frequencies are not provided for in
standard tomography, and if analyzed at either frequency, one of them
is incorrectly represented \citep{hellier99}.  As noted in
Sect.~\ref{sec:decom2}, the trailed spectra of the spin component
contain no sinusoidal structure or part thereof that the tomogram can
assign to a specific position in velocity space. No information is
obtained from these tomograms and we do not show them.

The orbital tomography of the narrow emission lines of \fei\ and \caii\ in
Fig.~\ref{fig:tomosec} proves their origin from the heated face of the
secondary star. Our measurement of the velocity amplitude of the
secondary star, $K_2=432.4\pm4.8$\,\kms\ (Paper~I, BR08), rested on
the interpretation of the \caii\ line profiles in terms of an advanced
illumination model.

\subsubsection{Orbital tomograms of the observed trailed spectra}
\label{sec:tomodisk}

The orbital tomograms of the observed spectra of \hbet\ and \hgam\ in
Fig.~\ref{fig:gray2} are shown in the two left panels of
Fig.~\ref{fig:tomo}. They compare well with the results of
\citet{belleetal05}, \citet{mhlahloetal07}, and
\citet{echevarriaetal16}, except that the emission of the irradiated
secondary star is not seen in their data. These tomograms include the
azimuthally smeared out spin component at high velocities, which
merges with the central representation of the disk features, masking
the transition between the two.

The tomogram of \ion{He}{i}$\lambda 4471$ is shown in the upper right 
panel of Fig.~\ref{fig:tomo}. As \hbet, it extends to high
velocities. Part of the enhanced emission around velocities of 1200
\kms\ is an artifact produced by the \ion{Mg}{ii}$\lambda4481$ line
that creates a false impression of the transition between the disk and spin
components.

The \heii\ spectra are single peaked and the intensity of the tomogram
is centered on the WD. The \heii\ tomogram in the lower right
panel of Fig.~\ref{fig:tomo} is representative of the emission in the
inner part of the accretion curtains seemingly with little
interference by emission   or absorption by matter in the disk. The
radial profile of the emission may be affected, however, by the
varying degree of helium ionization. Many details of the complex
radiative transfer in the magnetosphere and disk bulge remain to be
studied \citep{ferrarioetal93,kimbeuermann96}. A peculiar feature in
the \heii\ tomogram is an emission peak in the vicinity of $L_1$ that
could arise from matter lingering in front of $L_1$. Stellar
prominences, as observed in QS~Vir, provide a possible explanation
\citep{parsonsetal11}. Artifacts created by \ion{He}{i}$\lambda 4713$
are the long whitish arc in the lower left and the emission patch
around $\upsilon_\mathrm{x},\upsilon_\mathrm{y}\simeq+500,-1000$\,\kms.

\begin{figure*}[t]
\includegraphics[height=61mm,angle=270,clip]{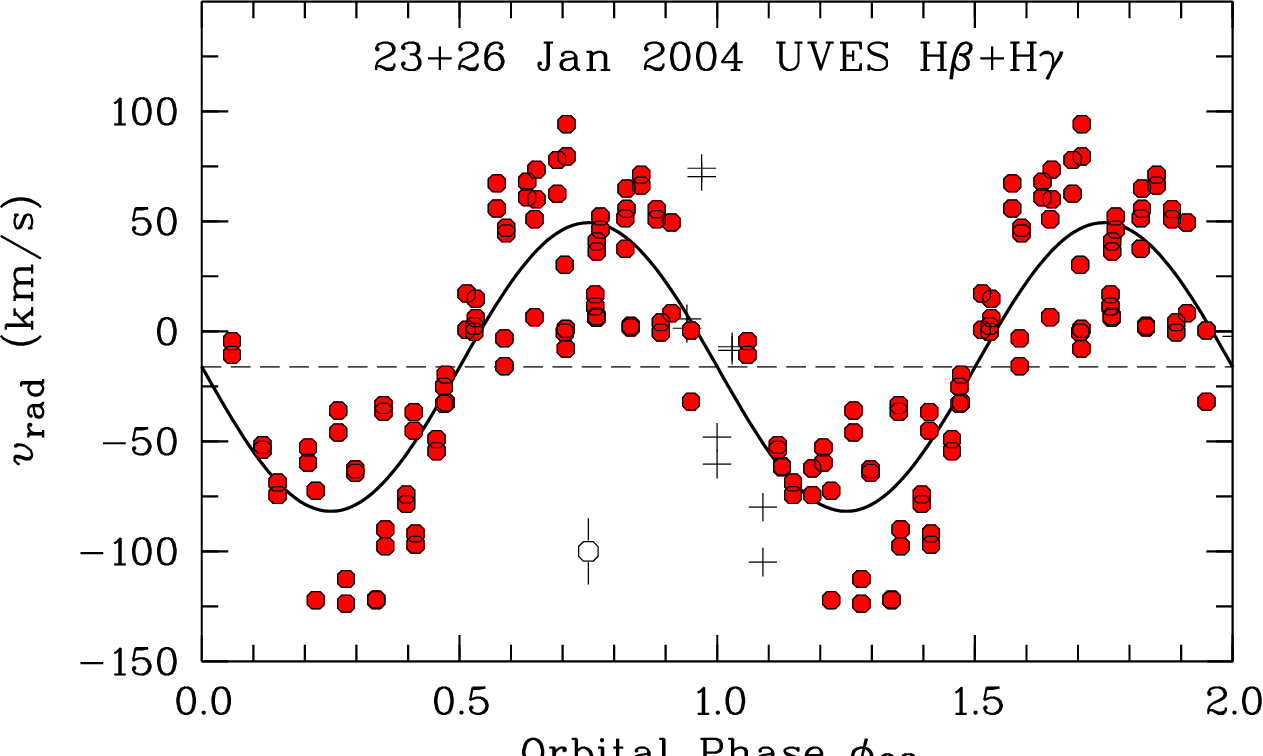}
\includegraphics[height=61mm,angle=270,clip]{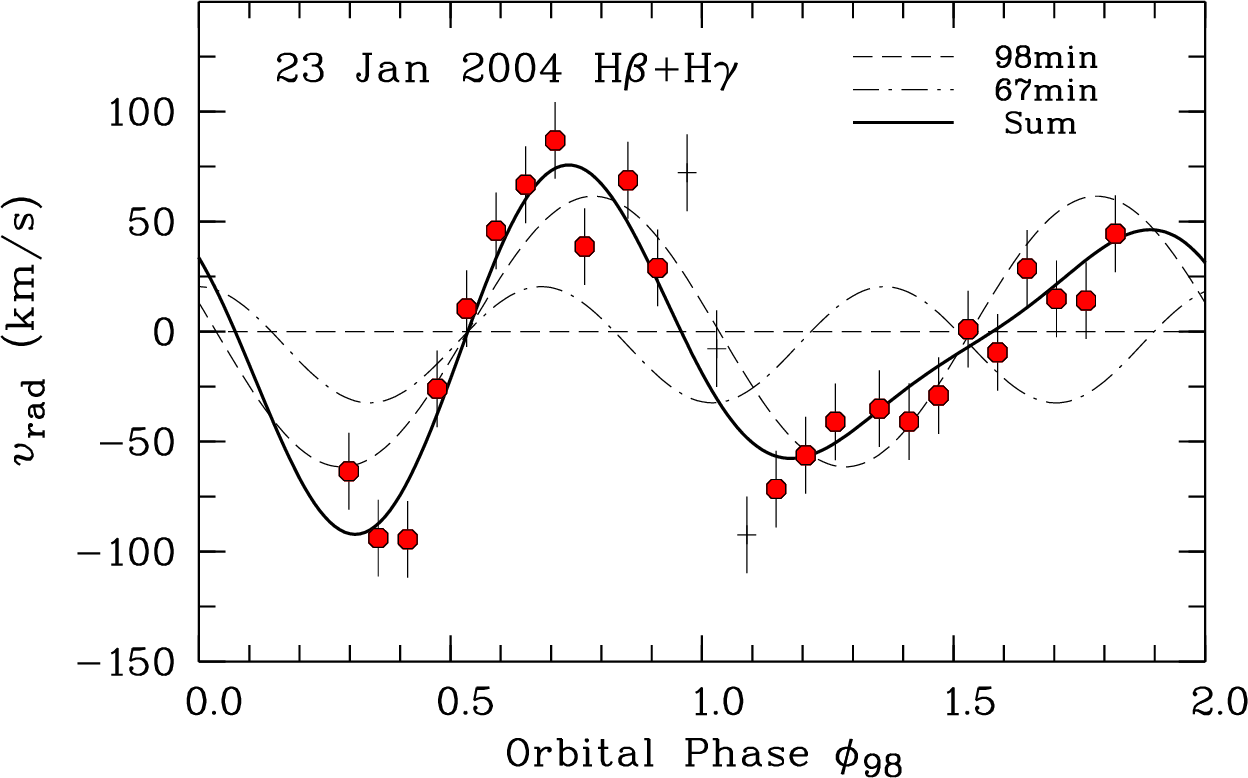}
\includegraphics[height=60.5mm,angle=270,clip]{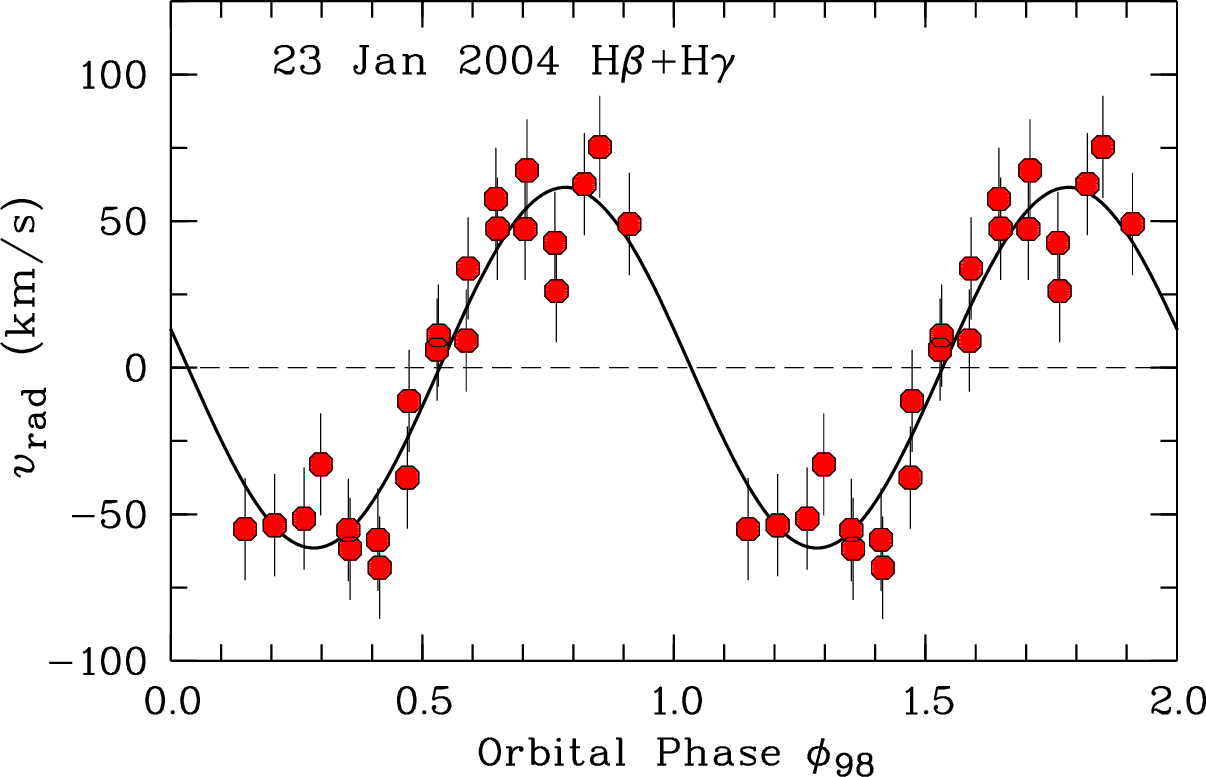}

\vspace{1mm}
\includegraphics[height=61mm,angle=270,clip]{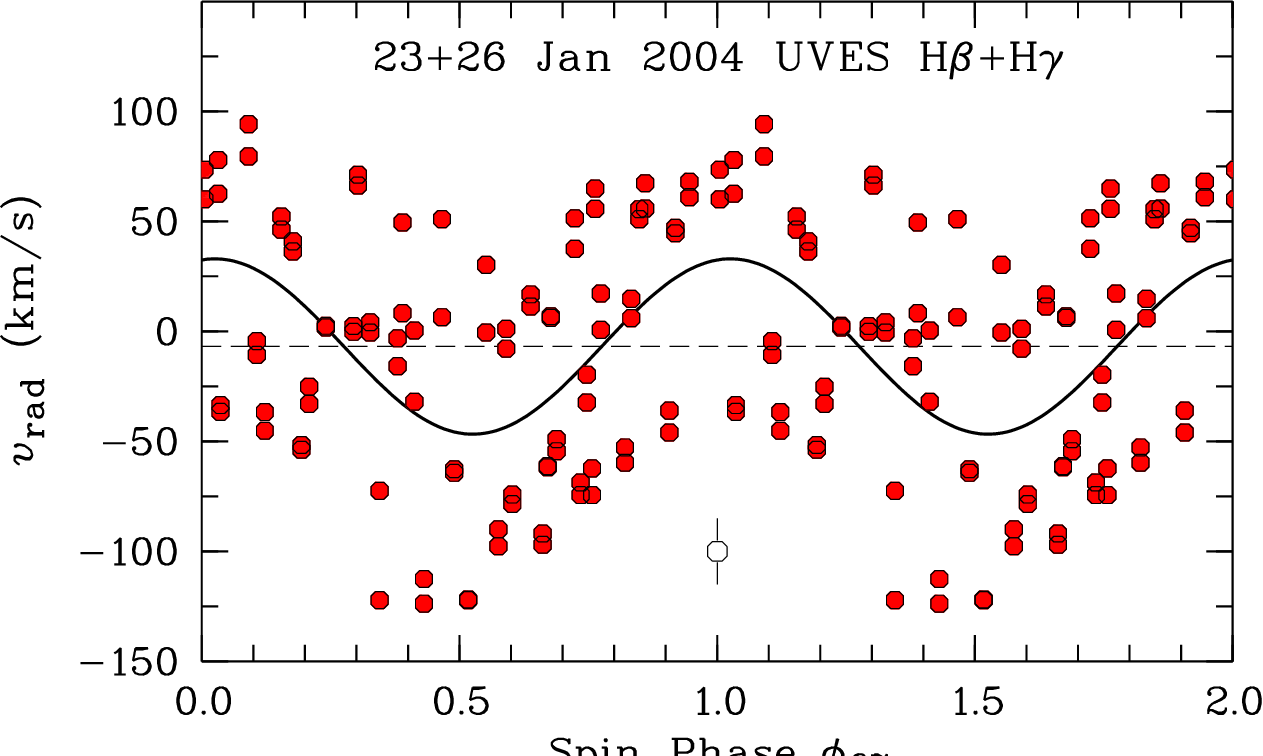}
\includegraphics[height=61mm,angle=270,clip]{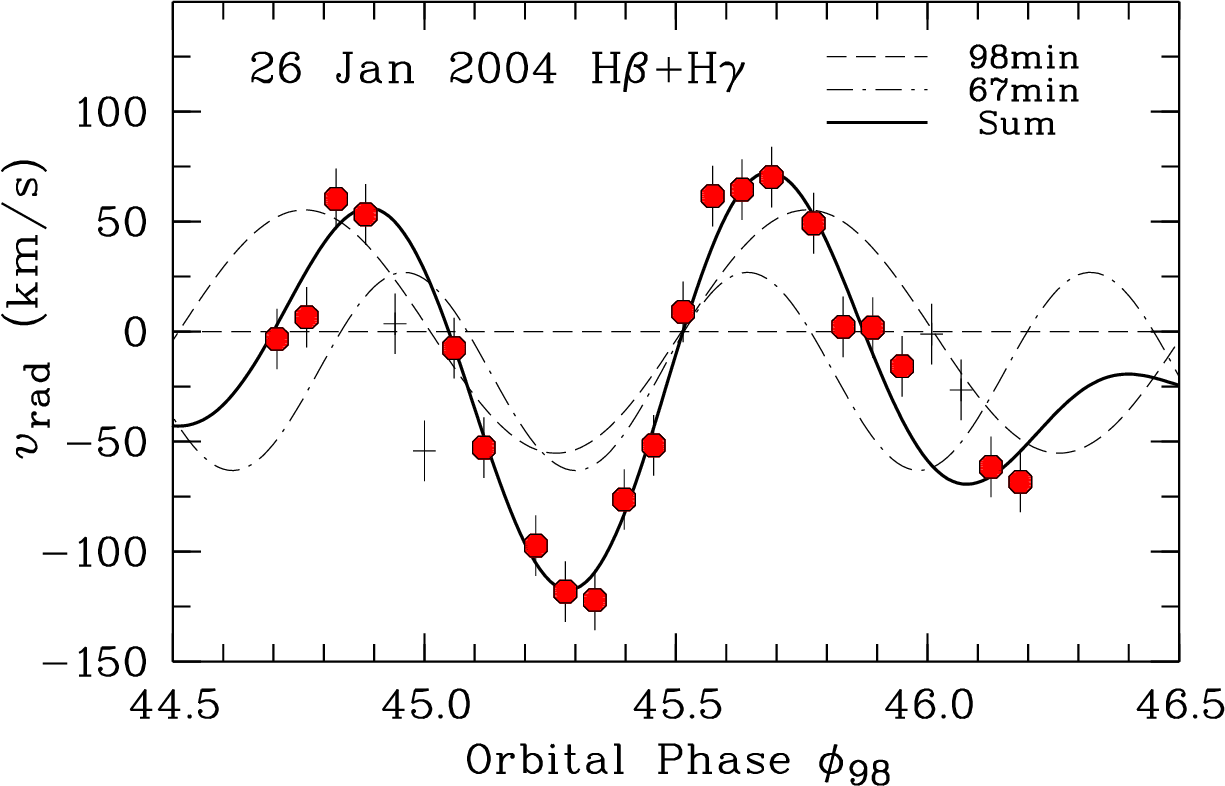}
\includegraphics[height=61mm,angle=270,clip]{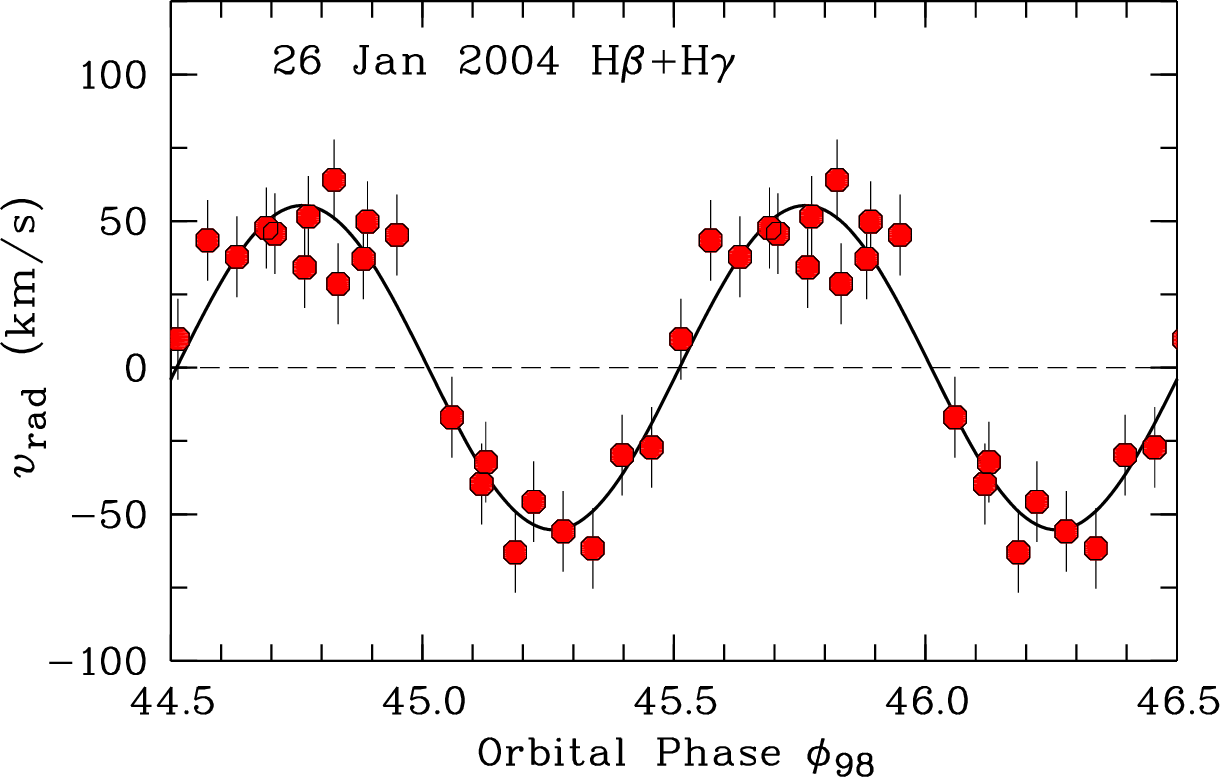}
\caption{
  Radial velocities of \hbet\ and \hgam\ for the nights of 23 and 26
  January 2004. Left panels: Individual velocities of \hbet\ and
  \hgam\ for both nights plotted vs. orbital phase (top) and vs. spin
  phase (bottom). A typical error is shown attached to the open
  circle. The solid curves represent single sine fits. Center panels:
  Mean radial velocities of \hbet\ and \hgam\ for 23 January (top) and
  26 January (bottom), fitted by Eq.~\ref{eq:9867} (solid curve). The
  dashed and dash-dotted curves indicate the components that vary on
  the orbital and the spin periods, respectively. Right panels: Same, but
  with the spin-modulated component subtracted.}
\label{fig:vrad}
\end{figure*}

\subsubsection{Tomography of the disk component}
\label{sec:tom}

The tomograms of the core components of \hbet\ and \hgam\ are
displayed in the two central panels of Fig.~\ref{fig:tomo}.  Their
most obvious feature is (i) the strict limitation to velocities below
about 1200\,\kms. Additional features include (ii) an asymmetry of the
terminal velocity of the disk; (iii) the emission from the ballistic
stream and the hot spot; (iv) additional emission from farther
downstream, possibly indicating stream overflow, and (v) the emission
from the irradiated face of the secondary star. We deal with these
topics in turn.

Ad (i): The double-peaked Balmer line profile is mapped into the broad
ring of emission that extends from about 500 to beyond 1100\,\kms,
vanishing at $1300$\,\kms.  The last two numbers define the range over
which the intensity in the tomogram decreases. Reducing the smoothing
parameter does not cause the transition to become significantly
sharper, suggesting that there is, in fact, such a transition
region. The quoted velocity range corresponds to distances from the WD
between $6.0\,\times\!10^9$ and $8.3\,\times\!10^9$\,cm or
$8.2\!-\!11.3$\,\reins\ and $64\!-\!88$\% of the circularization
radius.\footnote{We quote radii on the assumption of circular motion,
  although this is not exactly true given the observed asymmetry of
  the motion at the inner edge of the disk noted in Ad (ii).}  This
appears plausible in the diamagnetic blob model of \citet{king93},
which is characterized by non-circular and non-Keplerian motion in the
vicinity of \rcirc\ and involves a finite radial transition region
over which the individual blobs attach to the magnetic field of the
WD. For definiteness, we quote a mean terminal velocity in the
transition region of \vterm$\,\simeq\!1200$\,\kms, and an inner radius
of the disk \rin$\,\simeq\!\ten{7}{9}$\,cm or close to 10\reins,
effectively recovering the result of \citet{hellieretal87}.

Ad (ii): The Balmer line emission at the inner edge of the disk
is not exactly centered on the WD, but displaced by $\sim\!100$\,\kms\ toward
azimuth $\psi\!\simeq110$\degr, indicating a slightly non-circular
motion at the inner disk rim. Its presence and the observed phasing
are as expected from the diamagnetic blob model
\citep{king93,kingwynn99,kunzeetal01}.

Ad (iii): In all tomograms the bright spot is located at
$\upsilon_\mathrm{x},\upsilon_\mathrm{y}\!\simeq\!-500,+420$\,\kms,
above the single-particle trajectory, implying that the stream is
accelerated azimuthally as it interacts with the Keplerian disk
\citep[see also][]{mhlahloetal07,echevarriaetal16}.  SMP model
calculations by \citet{kunzeetal01} nicely illustrate this process.

Ad (iv): The main topic of \citet{kunzeetal01} was the ubiquity
of stream overflows in CV disks. Their calculations are helpful in
identifying the path of the stream and its re-impact on the disk.  In
our tomograms, bright regions that follow the hot spot in azimuth may
indicate the stream near peak altitude above the disk
($\upsilon_\mathrm{x},\upsilon_\mathrm{y}\simeq-1000,+0$\,\kms) and
the re-impact on the disk
($\upsilon_\mathrm{x},\upsilon_\mathrm{y}\simeq-100,-700$\,\kms).

Ad (v): One of the bright emission peaks in the \hbet\ tomogram
is the irradiated face of the secondary star.  In \hgam\ the secondary
is fainter and the hot spot brighter, indicating a steeper Balmer
decrement for the chromospheric emission from the secondary star than
for the hot spot.
The emission from the secondary star is faint in \hei$\lambda 4471$
and is not detectable in \heii$\lambda 4686$.

\subsection{Orbital velocity of the WD and system parameters}
\label{sec:system}

A reliable measurement of $K_1$ from the Balmer line wings requires
that the very high positive and negative radial velocities cancel out.
Model calculations by \citet{ferrarioetal93} showed that almost
complete compensation is achieved if the curtains at both poles are
permanently visible.  This requires a large inner hole of the disk, and
results in a net velocity amplitude as low as $\sim\!40$\,\kms.
The existence of velocity variations at the spin period was indicated
by the detection of $V/R$ variations \citep{hellieretal87,
  mhlahloetal07}, but they have not yet been included in the orbital
velocity fits.
  
Our approach differs from that of other authors in two respects: we represented the line wings by appropriate templates rather than
Gaussians or Lorentzians, and  we accounted explicitly for radial
velocity variations at both \phiorb\ and \phispin.  In the left
panels of Fig.~\ref{fig:vrad} we show the individual radial
velocities of \hbet\ and \hgam\ of both nights folded over the orbital
(top) and the spin period (bottom). A clear modulation exists at both
periods, with the large scatter caused by the mutually competing
period.  The amplitudes are about 60\,\kms\ for the orbital and
40\,\kms\ for the spin modulation;  the latter is as predicted by
\citet{ferrarioetal93}. There is no evidence for variations at \porb/2
or \pspin/2 \citep{gilliland82,hellieretal87}. A much clearer picture
is obtained by fitting the mean radial velocities of \hbet\ and \hgam\
for each night as a function of time by the sum of two sinusoids varying with $\phi_{98}$
and $\phi_{67}$,
\begin{equation}
  \upsilon_\mathrm{rad} = K_1\,\mathrm{sin}[2\pi(\phi_{98}-\phi_1)]+K_\mathrm{S}\,\mathrm{cos}[2\pi(\phi_{67}-\phi_\mathrm{S})]+\gamma,
\label{eq:9867}
\end{equation}
\noindent
with amplitudes $K_1$ and $K_\mathrm{S}$, zero crossings $\phi_1$ and
$\phi_\mathrm{S}$, and a common $\gamma$ velocity (center panels of
Fig.~\ref{fig:vrad}).  Data points between \phiorb$\,\simeq\!0.95$ and
1.05 (crosses) were excluded from the fit, because they were
shown to be affected by the partial eclipse of the emission region
\citep{hellieretal87,echevarriaetal16}. As expected when two
periodicities contribute to the measurements, the combined
radial-velocity curves are not sinusoidal and differ for the two
nights. To keep the phases over the 45 orbits between January 23 and 26,
we chose the continually progressing phase \phiorb\ as an independent
variable. The fitted orbital and spin components are individually
shown by the dashed and dot-dashed curves, respectively, the sum by
the fat solid curve. Subtracting the fitted spin component yields the
net orbital variations in the right-hand panels of
Fig.~\ref{fig:vrad}. The fit parameters are listed in
Table~\ref{tab:9867}. The $K_1$ values of the two nights agree within
their uncertainties. The weighted mean is $K_1=58.7\pm3.9$\,\kms.  The
different results for $K_\mathrm{S}$ may indicate that the ansatz of
Eq.~\ref{eq:9867} still leaves some aspects unaccounted for.

There is good agreement between the orbital radial velocity amplitudes
$K_1$ of the WD measured from X-ray and far-ultraviolet emission lines
that originate on or near the WD, and from the wings of the Balmer
emission lines. The first method gave $K_1\!=\!58.2\!\pm\!3.7$\,\kms
\citep{hoogerwerfetal04} and $59.6\!\pm\!2.6$\,\kms
\citep{belleetal03}, the second $58\,\pm\,5$\,\kms\
\citep{echevarriaetal16}, $58.7\!\pm\!3.9$\,\kms\ (this work), and
similar values in earlier publications (see Table~4 in Echevarria et
al. 2016). A weighted mean of the quoted four best-defined amplitudes
is $K_1=58.9\pm1.8$\,\kms.  Together with the velocity amplitude of
the secondary star $K_2\!=\!432.4\!\pm\!4.8$\,\kms\ and an inclination
$i\!=\!77\fdg8\!\pm\!0\fdg4$ (BR08), the component masses become
$M_1\!=\!0.790\!\pm\!0.034$\,\msun\ and $M_2\!=\!0.108\!\pm\!0.007$,
practically the same as in BR08 and in \citet{echevarriaetal16}.

\begin{table}[b]
  \caption{Parameters of a two-component fit to the mean radial
    velocities of \hbet\ and \hgam. $K_1$ is the amplitude of the
    orbital motion, $K_\mathrm{S}$ that of the spin modulation.}
\begin{tabular}{l@{\hspace{2mm}}c@{\hspace{2.5mm}}c@{\hspace{2.5mm}}c@{\hspace{3.0mm}}c@{\hspace{1.5mm}}c}
\hline \hline \noalign{\smallskip}
Date in &  $K_1$        & $\phi_1$      & $K_\mathrm{S}$ & $\phi_\mathrm{S}$ & $\gamma$      \\
  2004  &  \kms\        &               &    \kms\      &                  &   \kms\       \\[0.5ex]
  \noalign{\smallskip} \hline
  \noalign{\medskip}                                                            
23 Jan & $61.5\!\pm\!5.3$ & $0.53\!\pm\!0.02$ & $26.4\!\pm\!5.6$  & $0.05\!\pm\!0.02$   & \hspace{1.6mm}$-6.0\pm4.0$ \\
26 Jan & $55.3\!\pm\!5.8$ & $0.51\!\pm\!0.02$&$45.0\!\pm\!5.6$&\hspace{-1.7mm}$-0.04\!\pm\!0.02$  & $-18.4\pm4.0$ \\[0.5ex]
Mean   & $58.7\!\pm\!3.9$ & $0.52\!\pm\!0.02$ &                   & $0.01\!\pm\!0.02$   &               \\[0.5ex]
\noalign{\medskip} \hline
\end{tabular}
\label{tab:9867}
\end{table}

\subsection{Spin-phase dependence of the eclipse times}
\label{sec:ecl} 
Further information that helps to localize the inner disk edge can be
gathered from shifts of the optical and X-ray mid-eclipse times in
orbital phase. \citet{jablonskibusko85} and \citet{siegeletal89} found
that the optical timings shift back and forth by about $\pm20$\,s as a
function of spin phase, corresponding to a lateral displacement of the
centroid of the eclipsed light in the orbital plane of
$\pm10^9$\,cm. \citet{siegeletal89} located the emission in the
pre-shock region of the magnetically guided accretion stream, while
\citet{revnivtsevetal11} and \citet[][and refeences
therein]{suleimanovetal16,suleimanovetal19} placed it at the inner
edge of the accretion disk. Other authors assigned the eclipsed light,
at least in part, to the hot spot at the outer edge of the disk. Given
that the emission is probably distributed in the projected lateral
distance from the WD, a more detailed discussion is in place.

\begin{figure}[t]
  \includegraphics[width=41.0mm,angle=270,clip]{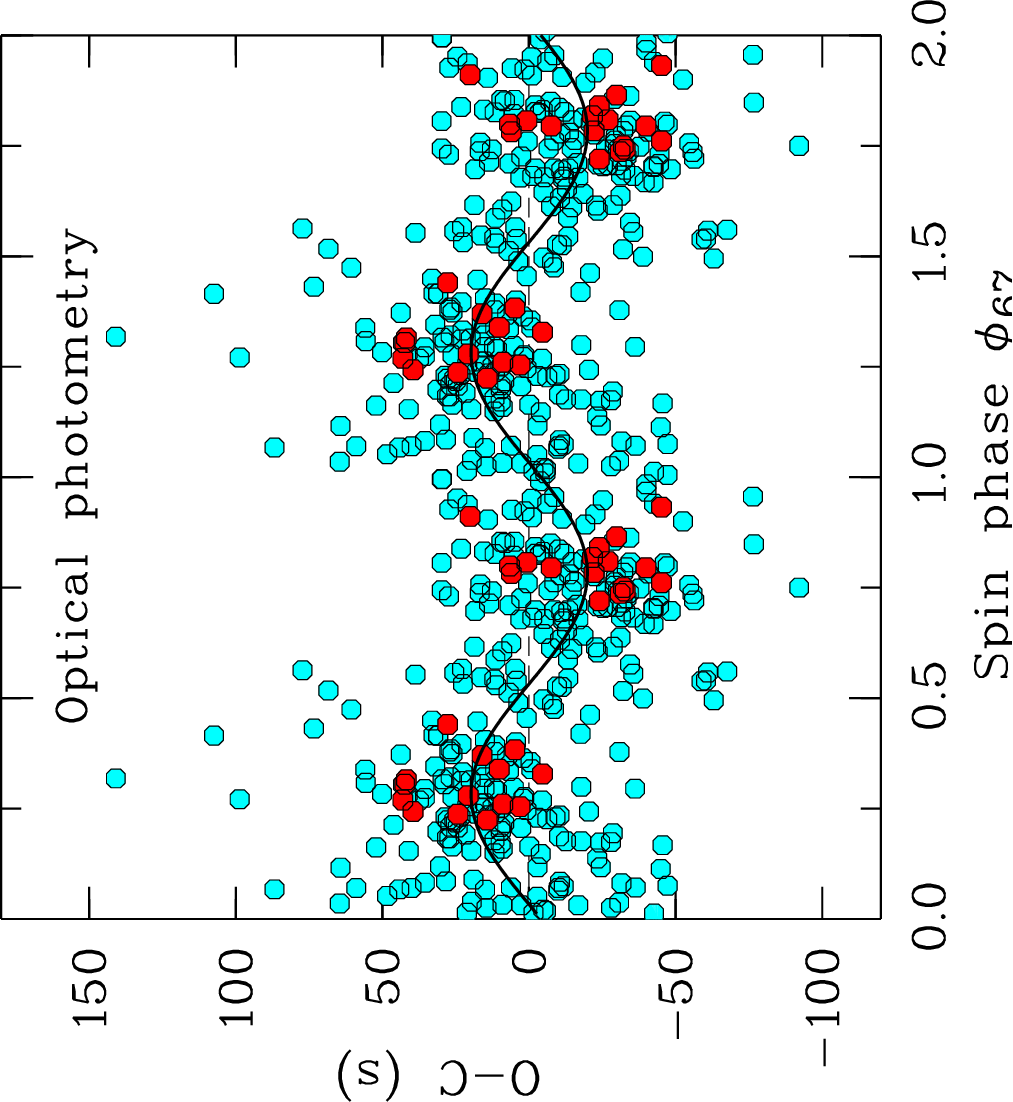}
  \hfill
  \includegraphics[width=41.0mm,angle=270,clip]{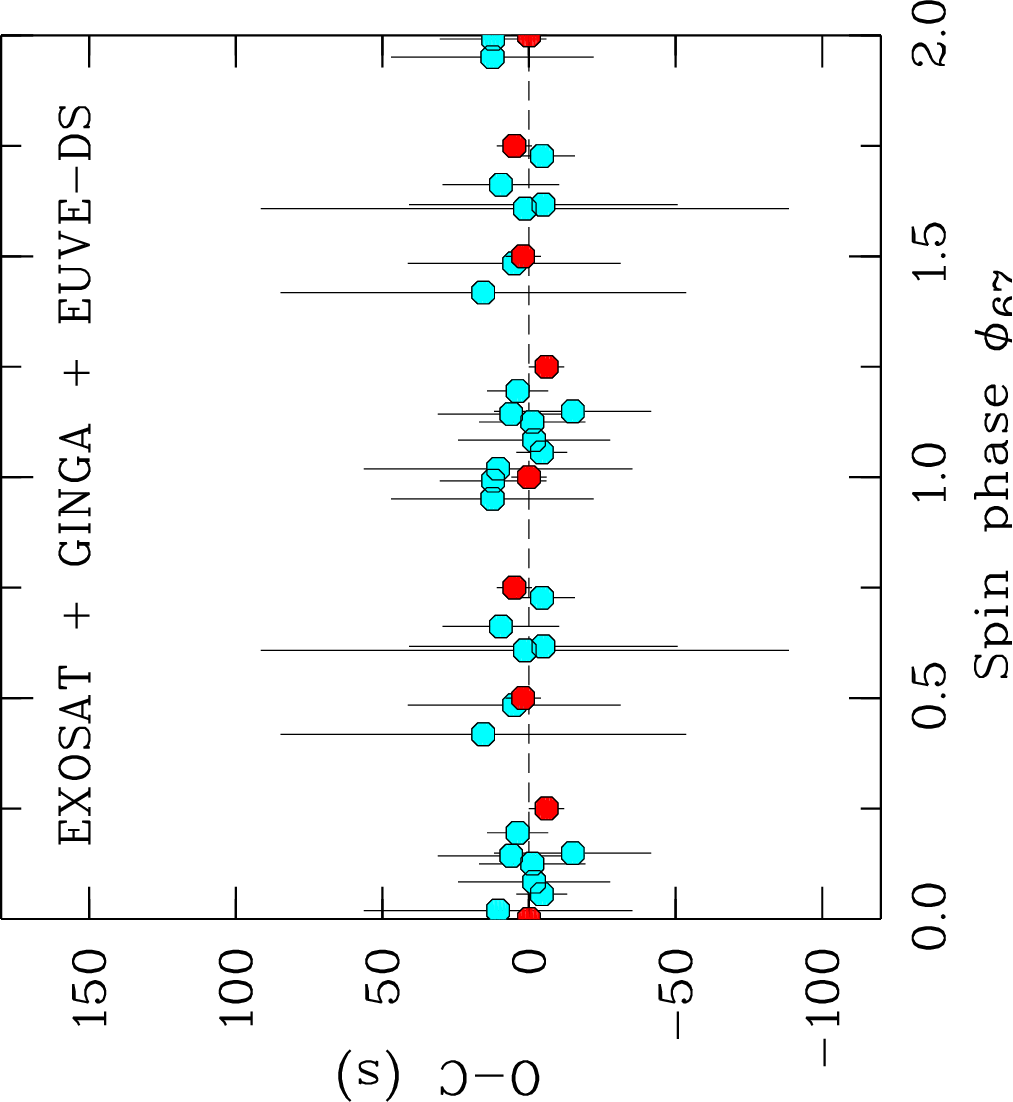}
  \caption{
    \oc\ deviations of the mid-eclipse times vs. spin phase. Left:
    Optical eclipse times collected by Echevarria et al. (2016, cyan)
    and by Siegel et al. (1989, red). The sinusoid fitted to all
    data has an amplitude of 20\,s. Right: X-ray and EUV
    eclipse times observed with EXOSAT and GINGA (Rosen et
    al. 1988,1991, cyan dots) and with the EUVE Deep Survey instrument
    (Hurwitz et al. 1997, Belle et al. 2002, red dots).}
\label{fig:spinphase}
\vspace{-3mm}  
\end{figure}

\begin{figure}[t]
  \includegraphics[width=89.0mm,angle=0,clip]{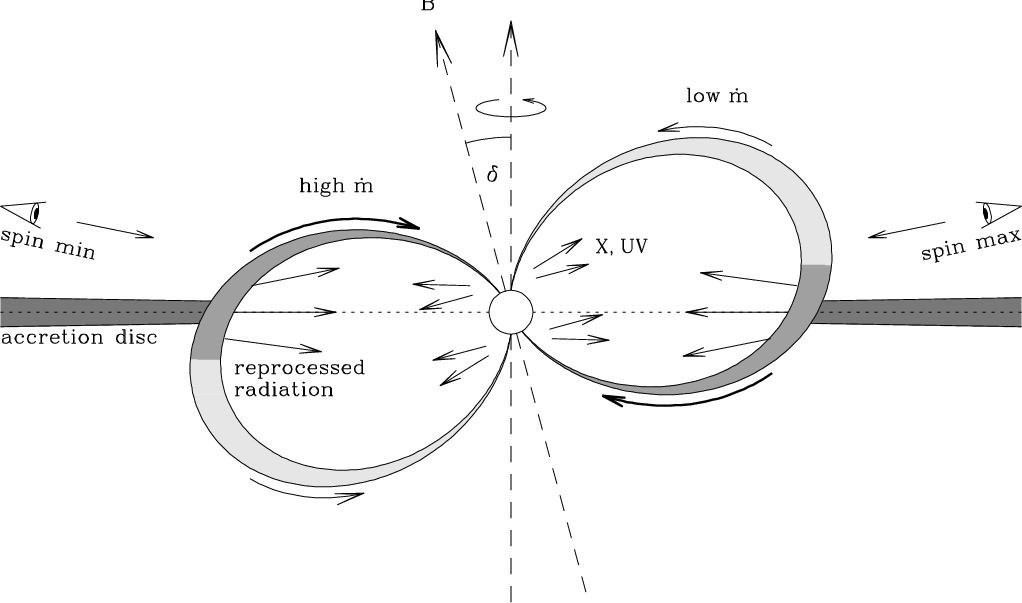}
  \caption{Schematic diagram of the magnetosphere and accretion region
    as viewed by the observer at spin phase 0.25. The different shades of
    gray indicate  different levels of the mass flow rate $\dot
    m$. Spin maximum occurred 0.25 rotations earlier at \phispin=0
    when the upper pole was pointing away from the observer and the
    lower accretion curtain was most directly in view.  }
\label{fig:model}
\vspace{-3mm}  
\end{figure}

In addition to the 31 mid-eclipse times of \citet{siegeletal89}, we made use
of the summary of 342 timings compiled by \citet{echevarriaetal16}
that contains data between 1962 and 2008 in seven subsets. Of these,
five pre-1986 data sets offer sufficient spin-phase coverage to allow
the necessary detrending for long-term variations of the eclipse times
of unknown origin \citep{jablonskibusko85,helliersproats92}. To this
end, we subtracted the mean $O-C$ for each subset.
The left panel of Fig.~\ref{fig:spinphase} shows the spin-phase
dependence of the detrended optical \oc\ with Echevarria's data as
cyan dots and Siegel's s data as red dots. A sine fit gives an
amplitude $\Delta t=20.0\pm1.9$\,s for the combined data, with maximum
positive (negative) \oc\ at \phispin$\,=\!0.25$ (0.75) (solid
curve). Individual fits to the six subsets show no significant
variation in amplitude or phase between 1962 and 1985. The information
for times after 1985 is too meager to allow a conclusion on the
long-term behavior of this modulation.

The right panel of Fig.~\ref{fig:spinphase} shows the spin-phase
dependence of the extreme ultraviolet (EUV) and X-ray mid-eclipse
times, which is based on more scanty data.  The eclipse times measured
in 1994 and 2000 with the Deep Survey Instrument of the Extreme
Ultraviolet Explorer (EUVE) are shown as red dots
\citep{hurwitzetal97,belleetal02} and the X-ray eclipse times observed
in 1985 with the European X-ray Observatory SATellite (EXOSAT) and in
1988 with the Japanese GINGA satellite as cyan dots
\citep{rosenetal88,rosenetal91}. We omitted three very ill-defined and
uncertain X-ray eclipse times near spin minimum. These mid-eclipse
times display no obvious dependence on spin phase, and we
conservatively estimate an amplitude $\Delta t\,<\!10$s.  We interpret
the different amplitudes for X-rays and optical light by their
respective origin below the strong shock near the WD and farther out
in the accretion curtain (see Fig.~\ref{fig:model}).

The observed amplitude $\Delta t$ reflects the lateral displacement of
the centroid of the eclipsed light from the accretion curtain in the
lower hemisphere \citep{beuermannosborne88},\footnote{The entire
  upper accretion curtain escapes eclipse.}
$\langle s_\mathrm{em}\rangle\!\simeq\!2\pi\,a\Delta t/P_\mathrm{orb}$
at \phispin\,=\,0.25 (Fig.~\ref{fig:model}). The true radial distance
$r_\mathrm{em}$ of the centroid of the eclipsed light may
substantially exceed $\langle s_\mathrm{em}\rangle$, depending on the
azimuthal distribution $I(\psi)$ of the emission.  With $\psi\!=\!0$
in the direction of the observer, the centroid of the emission
displays a relative shift
$\langle s_\mathrm{em}\rangle/r_\mathrm{em}=\langle \mathrm{sin}\psi
\rangle_{2\pi}$, where the right-hand term is the intensity-weighted
$2\pi$ average.  To give an example, we assume that accretion occurs
over $2\pi$ in azimuth and the emission varies as
$I(\psi)\!=\!1+a\,\mathrm{sin}\,\psi$, with $a\!<\!1$. Then, for
example, for $a\!=\!2/3$ the emission varies by a factor
$(1+a)/(1-a)\!=\!5$ between maximum (dark gray) and minimum (light
gray) and
$r_\mathrm{em}\!=\!(2/a)\langle s_\mathrm{em}\rangle\!=\!3\,\langle
s_\mathrm{em}\rangle$ or $r_\mathrm{em}\!\simeq\!\ten{3}{9}$\,cm for
$\Delta t\!=\!20$\,s. If the emission is centered halfway between
\rin\ and the WD, the example corresponds to
\rin$\,\simeq\ten{6}{9}$\,cm, about the value obtained from our
tomography.

\section{Discussion}
\label{sec:disc}

\subsection{Inner edge of the accretion disk}

We decomposed the \hbet\ and \hgam\ line profiles into a
spin-modulated component that is responsible for the entire line wings
and an orbital phase-modulated component that largely represents the
accretion disk. Our tomography shows the disk emission to be
restricted to \vrad$\,\la\!1200\,$\kms, tapering off between 1100 and
1300\,\kms, equivalent to Kepler radii between 6.0 and
$\ten{8.3}{9}$\,cm or $8.2$ and $11.3$\,\reins. In general, we confirm
the classical model that was established when \oex\ was recognized as
an IP \citep{kruszewskietal81,cordovaetal85,hellieretal87,kaitchucketal87}
and assigns the entire line wings to streaming motions in the
magnetosphere.

This model was challenged by \citet{revnivtsevetal09} who interpreted
a break in the power spectra of the optical and X-ray temporal
fluctuations with the Kepler frequency at the inner edge of the disk,
seemingly an elegant method of measuring \rin. Applications by
\citet{revnivtsevetal11}, \citet{semenaetal14},
\citet{suleimanovetal16}, and \citet{suleimanovetal19} gave inner disk
radii for \oex\ between 2 and 4\,\reins, implying that velocities in
the Balmer line wings up to between 2600 and 1800\,\kms\ represent
Keplerian motion in the disk and velocities beyond streaming motions
in the magnetosphere. This dichotomy is inconsistent with our finding
that the entire Balmer line wings at \vrad$\,\ga\!1200$\,\kms\ vary
with spin phase as a single component. Furthermore, \heii\ line
emission originates almost exclusively from the magnetospheric funnels,
and this will hold for strong far-ultraviolet emission lines as
\ion{Si}{iv}$\lambda 1400$ as well. It is unlikely that the Balmer
line wings behave so differently. A variety of optical and X-ray
observations demand that the accretion curtains on the WD in \oex\ be
tall and the lower one be visible through the hole in the inner
disk. With a required shock height \hsh$\,\sim\!1$\,\reins\
\citep{allanetal98,lunaetal18}, an inner disk radius of about
9\,\reins\ is required. If  this requirement is violated, the
small-magnetosphere model runs into problems, which  explains a wide range
of observations, including (i) the partial nature of the X-ray eclipse
\citep{beuermannosborne88,rosenetal91,mukaietal98}, (ii) the energy
dependence of the X-ray spin modulation \citep{rosenetal91}, (iii) the
observed shock temperature (Sect.~\ref{sec:tsh}), (iv) the weak Fe
K$\alpha$ 6.4\,keV line and the absence of a strong reflection
component in the hard X-ray spectrum \citep{lunaetal18}, and (v) the
almost complete compensation of the streaming motions from both
curtains \citep[][and this work]{ferrarioetal93}. Hence, there is
ample evidence for a large inner hole in the accretion disk of
\oex. We conclude that the model of \citet{revnivtsevetal09} yields an
incorrect result, and we agree with \citet{lunaetal18}, who considered its
basic assumption as suspect.

In many IPs at longer orbital periods, \rin$\,\simeq\,$\rco, which causes
them to accrete near spin equilibrium. The large-magnetosphere model
defines the conditions under which \oex\ can be considered to accrete
in spin equilibrium \citep{kingwynn99,nortonetal04,nortonetal08}.
\citet{belleetal02} and \citet{mhlahloetal07} discussed their data
within such a model, but did not quote a value of \rin. We
discuss this model in   Sect.~\ref{sec:moment}.

\subsection{Shock temperature}
\label{sec:tsh}

The freefall of accreting matter to the WD starts at \rin\ and is
braked at the shock at height \hsh\ above its surface.  The
electron temperature behind a strong shock is given by
\begin{equation}
\mathrm{k}T_\mathrm{sh}=34.3\,\left(\frac{1}{1+(h_\mathrm{sh}/R_1)}-\frac{1}{r_\mathrm{in}/R_1}\right),
\label{eq:tsh}
\end{equation}
where the numerical factor is
$\mathrm{k}T_\mathrm{sh}=(3/8)\,\mu
m_\mathrm{u}\mathrm{G}M_1/R_1=34.3$\,keV; $\mu=0.619$ is the mean
molecular weight for a plasma of solar composition; and $m_\mathrm{u}$
is the unit mass. We used \mwd$=0.79$\,\msun\ and
\reins$\,=\!\ten{7.35}{8}$\,cm (BR08). The published shock
temperatures were measured by fitting cooling flow models to the
observed hard X-ray spectra. Published values include 12.7\,keV
\citep{yuasaetal10}, 15.4\,keV \citep{fujimotoishida97}, 16\,keV
\citep{hoogerwerfetal04}, 18.0\,keV \citep{hayashiishida14}, 19.4\,keV
\citep{brunschweigeretal09}, 19.7\,keV \citep{lunaetal18}, and
$\sim\!20$\,keV \citep{mukaietal03}.
\noindent For \rin\,=\,10\reins, Eq.~\ref{eq:tsh} reproduces the
quoted range of the observed shock temperatures for shock heights of
$0.5-1.2\,R_1$, in agreement with the observationally required
\hsh$\,\sim\!1.0$.

\subsection{Magnetic moment of the white dwarf}
\label{sec:moment}

The magnetic moment $\mu_1$ of the WD is directly related to the size
of \rin. Obtaining an estimate of $\mu_1$ is   straightforward
if we assume that the theory of the interaction of a viscous disk with
a magnetosphere \citep[e.g.,][see also White\,\&\,Stella
1988]{ghoshlamb79} is applicable to \oex. In this theory, azimuthal
magnetic stresses bring the plasma into corotation with the field at a
radius
\begin{equation}
  r_\mathrm{mag}\,\simeq\,\zeta\,\mu_1^{4/7}(2GM_1)^{-1/7}\dot M^{-2/7} ~~~\mathrm{cm},
 \label{eq:rmag}
\end{equation}
where Kelvin-Helmholtz instabilities cause it to be invaded by the
stellar field and accreted \citep{aronslea80}.
Here $M_1$ is the mass of the WD, $\mu_1\!=\!B_1R_1^3$ its magnetic
moment, $B_1$ the surface field strength, $\dot M$ the accretion rate,
and $\zeta$ a factor that depends on, among other parameters,  the ratio
of the angular velocities of field and plasma (fastness parameter).
For the parameters of \oex, $\zeta\!\simeq\!0.45$ \citep[see
also][]{whitestella88}. Equating $r_\mathrm{mag}$ with
\rin$\,\simeq\!\ten{7}{9}$\,cm and using $M_1\!=\!0.79$\,\msun\ and
$\dot M\!=\!\ten{3.9}{-11}$\,\msunyr, gives
$\mu_1\!\simeq\!\ten{1.3}{32}$\,Gcm$^3$ and $B_1\!\simeq\!0.35$\,MG.
With a polar field strength of less than $1$\,MG, no optical circular
polarization is expected.  The quoted $\dot M$ is based on the overall
spectral energy distribution of \citet{eisenbartetal02} that we
updated and converted to spin maximum. Its integrated flux,
$F_\mathrm{acc}\!=\!\ten{8.2}{-10}$\,\ergs, represents our best
estimate of the $4\pi$-averaged emission of \oex,\ and leads to the
quoted $\dot M$ at $d=56.77$\,pc \citep{bailerjonesetal21}. The result
agrees closely with the mass transfer rate expected from gravitational
radiation and the spin-up of the WD \citep{ritter85}.

It is instructive to consider the condition under which \oex\ might be
able to synchronize. It requires that the magnetic torque
$G_\mathrm{mag}\,=\!\gamma\,\mu_1\,\mu_2\,/a^3$\,\gcms\ exceeds the
accretion torque
$G_\mathrm{acc}\,=\!\dot M(\mathrm{G}M_1r_\mathrm{mag})^{1/2}$\,\gcms,
where $\mu_2\,=\!B_2\,R_2^3$ is the magnetic moment of the secondary
star, $B_2$ and $R_2$ are its field strength and radius, and $\gamma$ is a
geometry-dependent factor of order unity \citep{kingetal90}. The
secondary in \oex\ is taken to possess a dynamo similar to the rapidly
rotating field stars studied by \citet{reinersetal09}. Stars with
Rossby number $Ro\!<\!0.10$ possess a saturated dynamo that generates
a surface field of roughly 3\,kG with a scatter of 1\,kG. The stars
with the lowest $Ro\!\sim\!0.01$ have field strengths of around 2\,kG,
which we adopt tentatively as the field strength of the secondary in
\oex. With $R_2\,=\!\ten{1.055}{10}$\,cm (BR08), the secondary in
\oex\ would then have a magnetic moment
$\mu_2\!\sim\!\ten{2.3}{33}$\,\gc\ and the magnetic torque becomes
$G_\mathrm{mag}\!=\!\ten{2.9}{33}\gamma$\,\,\gcms, while the accretion
torque is $G_\mathrm{acc}\!=\!\ten{2.1}{33}$\,\gcms. Hence, the
torques are of similar magnitude, and one may speculate that \oex\ was
in synchronous rotation in the past, but failed to regain it, for example
when $\dot M$ changed. We recall that the disk emission vanishes at
$r\!=\!\ten{6}{9}$\,cm, slightly exceeding the radius of closest
approach to the WD in ballistic orbits \citep{lubowshu75}, suggesting
that formation of the disk may have been marginal.

In the alternative picture of \citet{king93} and \citet{kingwynn99},
the disk in \oex\ consists of diamagnetic blobs that interact
individually with the local field through a surface drag, thereby
exchanging orbital energy and angular momentum with the stellar
components.  The interaction proceeds via the emission of Alfv\'en
waves generated as the blobs pass across field lines, and ``pluck them
as violin strings'' \citep{drelletal65,king93}. The timescale of this
process, \tmag$\,\simeq\!c_\mathrm{A}\rho_\mathrm{b}l_\mathrm{b}/B^2$,
depends on the local field strength, on the Alfv\'en velocity
$c_\mathrm{A}$ in the inter-blob medium, and on the blob density and
size $\rho_\mathrm{b}$ and $l_\mathrm{b}$.  Complex flow patterns may
accrue if blobs with a wide parameter range are present \citep[][their
Fig.~4]{kingwynn99}.
An overview and a classification of the flow patterns was presented by
\citet{nortonetal04} and \citet{nortonetal08}. For an intermediate
range of WD magnetic moments the flow is either a stream or a
propeller, depending on the spin period. A disk-like or a ring-like
structure develops at the low or the high end of the $\mu_1$ scale,
respectively \citep[][their Figs. 1 and 2]{nortonetal08}. The ring-like
structure was proposed by \citet{kingwynn99} in an attempt to show
that conditions exist under which \oex\ may accrete in spin
equilibrium. A magnetic moment $\mu_1\!>\!10^{35}$\,\gc\ is required
for this possibility and a ring is formed at the WD Roche radius or,
equivalently, the corotation radius, from which the WD accretes over
360\degr\ in azimuth. We can exclude both the high $\mu_1$ and the
large \rin. The second possibility is the disk mode that is realized
for $\mu_1\!<\!\ten{3}{33}$\,\gc\ at the orbital period of \oex.
We conclude that \oex\ resides securely in this section of Norton's
zoo.  Further study is needed to shed more light on the evolutionary
status of \oex.

\section{Conclusions}

Our main results can be summarized as follows:\\
(1) We decomposed the Balmer line profiles of \oex\ in a unique way
into the spin-modulated wings and an orbital-phase modulated core. The
former represents the magnetically guided accretion stream; the latter
is dominated by the accretion disk.
\\
(2) Tomography of the core component shows a broad ring of emission
from the disk and S-wave component, both confined to
\vkep$\,\la\!1200$\,\kms\ or
\rin$\,\simeq\!7\!\times\!10^9\,$\,cm$\,\simeq10\,R_1\simeq0.75$\rcirc.
\\
(3) The inner hole of the disk is sufficiently large to permit an
unobstructed view of the lower accretion column for shock heights up
to \hsh$\,\simeq1.0\,R_1$, as required by a wide range of optical and
X-ray observations.
\\
(4) We estimated a WD magnetic moment
$\mu_1\!\simeq\!\ten{1.3}{32}$\,\gc\ and a surface field strength of
$0.35$\,MG.  With a polar field strength $B_\mathrm{p}\!\la\!1$\,MG,
no optical circular polarization is expected.
\\
(5) The measured \rin\ excludes the small-magnetosphere model of
\citet{revnivtsevetal09}, and the estimate of $\mu_1$ probably excludes
the large-magnetosphere model of \citet{kingwynn99}.  \oex\ is found
to accrete far from spin equilibrium.
\\
(6) The similar magnitudes of the magnetostatic and accretion torques
suggest conditions possibly close to synchronization.
\\
(7) The orbital velocity amplitude of the WD, $K_1\!=\!58.7\!\pm\!3.9$
\kms, agrees perfectly with previous results, confirming the published
component masses.

\begin{acknowledgements}
  We thank the anonymous referee for a quick, constructive, and
  helpful response. We thank Axel Schwope for providing the program
  for Doppler tomography originally written by Henk Spruit and for
  numerous enlightening discussions. We thank Stefan Eisenbart for
  Fig.~10 from his PhD Thesis (U of G\"ottingen, 2000).
\end{acknowledgements}

\bibliographystyle{aa}

\end{document}